\documentclass[preprint,number,3p]{elsarticle}

\usepackage{graphicx}
\usepackage{url}
\usepackage{subfig}

\begin{document}
\begin{frontmatter}{}

\title{Beam Position Reconstruction for the g2p Experiment in Hall A at Jefferson Lab}

\author[USTC]{Pengjia Zhu\corref{mycorresponding}}
\cortext[mycorresponding]{Corresponding author}
\ead{pzhu@jlab.org,zhupj55@mail.ustc.edu.cn}
\author[JLAB,MIT]{Kalyan Allada}
\author[JLAB]{Trent Allison}
\author[UNH]{Toby Badman}
\author[JLAB]{Alexandre Camsonne}
\author[JLAB]{Jian-ping Chen}
\author[WM]{Melissa Cummings}
\author[UVA]{Chao Gu}
\author[DUKE]{Min Huang}
\author[UVA]{Jie Liu}
\author[JLAB]{John Musson}
\author[UNH]{Karl Slifer}
\author[UVA,MIT]{Vincent Sulkosky}
\author[USTC]{Yunxiu Ye}
\author[JLAB,UVA]{Jixie Zhang}
\author[UNH]{Ryan Zielinski}

\address[USTC]{University of Science and Technology of China,Hefei,Anhui 230026,China}
\address[JLAB]{Thomas Jefferson National Accelerator Facility,Newport News,VA 23606,USA}
\address[UNH]{University of New Hampshire,Durham,NH 03824,USA}
\address[WM]{College of William \& Mary,Williamsburg,VA 23187,USA}
\address[UVA]{University of Virginia,Charlottesville,VA 22904,USA}
\address[DUKE]{Duke University,Durham,NC 27708,USA}
\address[MIT]{Massachusetts Institute of Technology,MA,02139,USA}

\begin{abstract}
Beam-line equipment was upgraded for experiment E08-027 (g2p) in Hall A at Jefferson Lab. Two beam position monitors (BPMs) were necessary to measure the beam position and angle at the target. A new BPM receiver was designed and built to handle the low beam currents (50-100 nA) used for this experiment. Two new super-harps were installed for calibrating the BPMs. In addition to the existing fast raster system, a slow raster system was installed. Before and during the experiment, these new devices were tested and debugged, and their performance was also evaluated. In order to achieve the required accuracy (1-2 mm in position and 1-2 mrad in angle at the target location), the data of the BPMs and harps were carefully analyzed, as well as reconstructing the beam position and angle event by event at the target location. The calculated beam position will be used in the data analysis to accurately determine the kinematics for each event.\end{abstract}
\end{frontmatter}{}

\section{Introduction}

A polarized ammonia ($NH_{3}$) target was used for the first time in Hall A for the g2p experiment \citep{g2pproposal}. It operated at a low temperature of 1K and a strong transverse magnetic field of either 5 T or 2.5 T. A high electron beam current would cause significant target polarization drop due to target temperature rising and ionization radiation to the target material \citep{Crabb67}. To minimize depolarization, the beam current was limited to below 100 nA and a raster system was used to spread the beam spot out to a larger area. The transverse magnetic field in the target region would cause the beam to be deflected downward when the beam enters the target region. To compensate for this, two chicane magnets were placed in front of the target to pre-bend the beam upwards. Due to the low beam current and tight space limitations after the chicane magnets, the experimental accuracy goals for the position (1-2 mm) and angle (1-2 mrad) at the target were challenging to achieve. New beam-line devices and an associated readout electronics system were designed for the g2p experiment to accomplish these goals. Design details and the performance of the beam-line devices will be described in the following sections along with a discussion of an analysis method determine the beam position and direction. 
\begin{figure}[tbph]
\begin{centering}
\includegraphics[width=1\columnwidth]{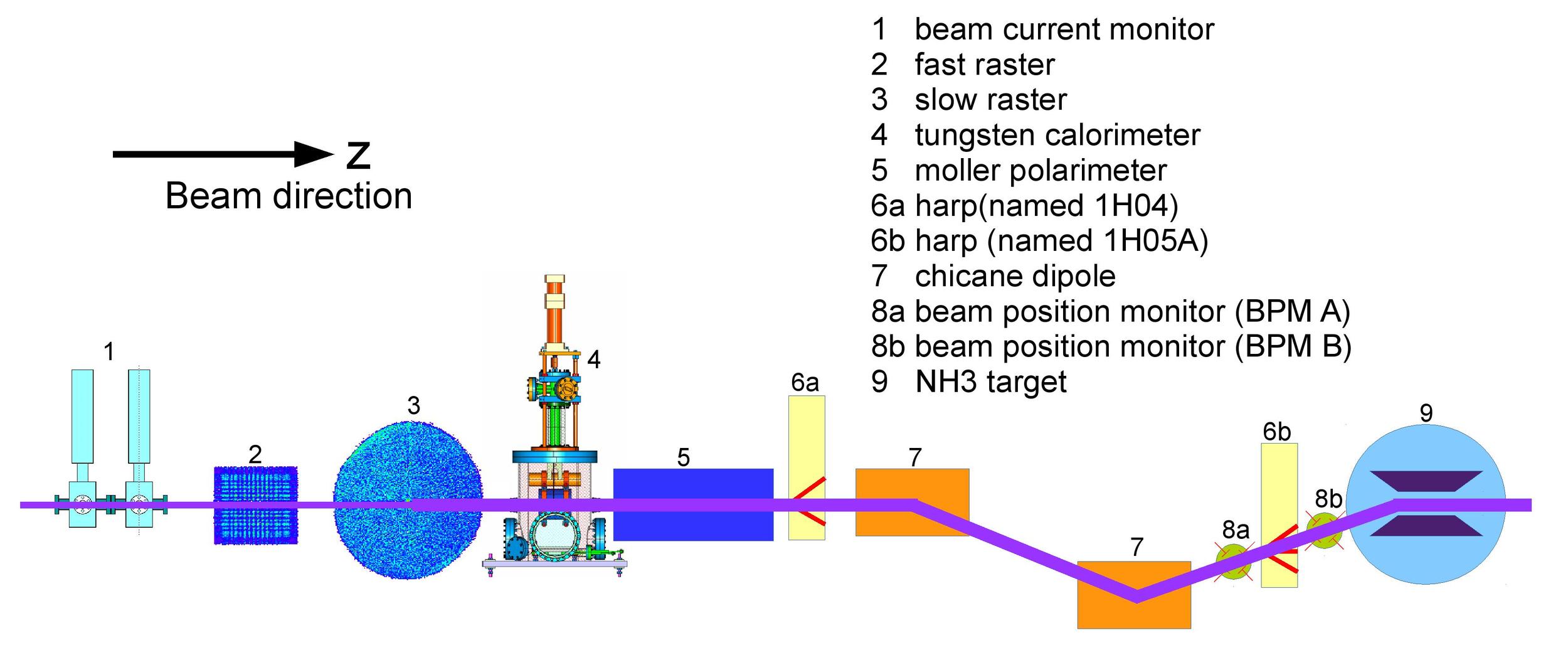} 
\par\end{centering}

\protect\caption{\label{fig:beamline-for-g2p}Schematic of beamline components for g2p experiment}
\end{figure}

\section{Beam-line Instrumentation}

\subsection{Beam position monitor (BPM)}

The scattering angle of the outgoing lepton in deep inelastic scattering, which is defined with respect to the direction of the incident beam, is an important variable for obtaining meaningful physics results. Therefore, the position and direction of the beam, after being bent by the chicane magnetic field and spread out by the rasters, must be measured precisely. Two BPMs and two harps were installed for relative and absolute measurements of beam position and direction near the target, respectively.

\begin{figure}[tbph]
\begin{centering}
\subfloat[BPM design diagram, from JLab instrumentation group]{\protect\begin{centering}
\protect\includegraphics[width=0.4\columnwidth]{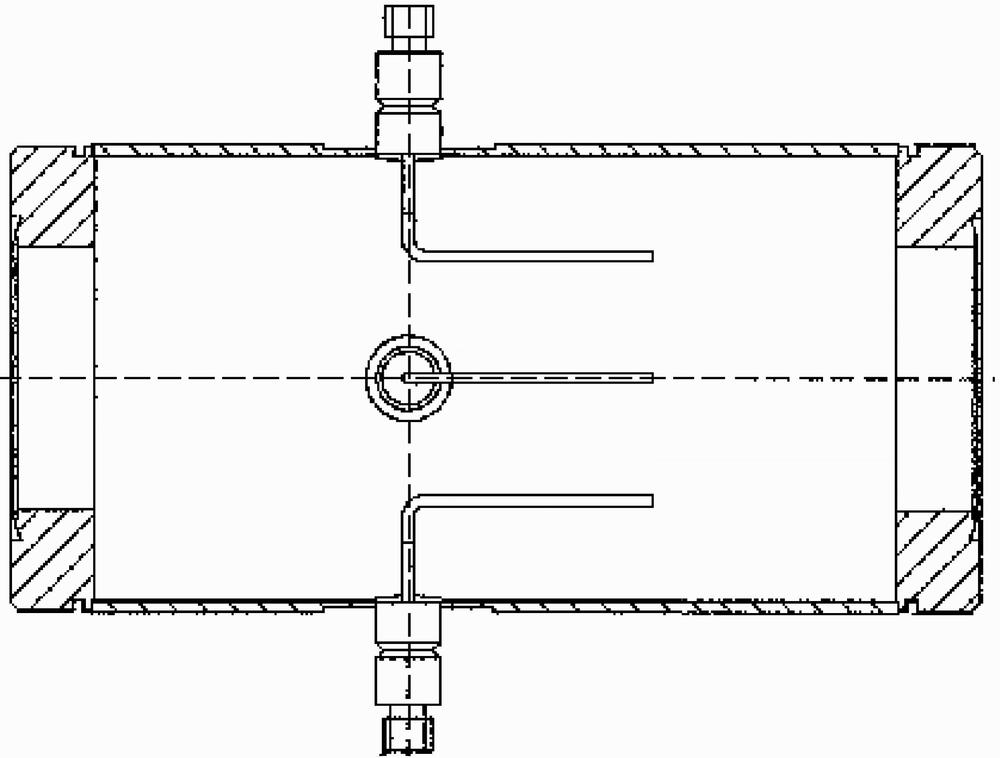}\protect
\par\end{centering}

}$\qquad$\subfloat[BPM chamber which contains 4 antennas]{\protect\begin{centering}
\protect\includegraphics[width=0.4\columnwidth]{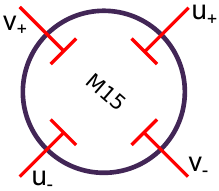}\protect
\par\end{centering}

}
\par\end{centering}

\protect\caption{\label{fig:bpm-design-diagram}BPM chamber}
\end{figure}

The BPM consists of four open-ended antennas for detecting the beam position; the measurement is non-invasive to the beam. The BPM chambers shown in Fig.\ref{fig:bpm-design-diagram} are part of the beam pipe. The four antennas are attached to feedthroughs on the interior wall of the pipe at $90^{\circ}$ intervals. 
 The BPM chambers are placed with a $45^{\circ}$ rotation (along z) with respect to the global Hall coordinate. The two pairs of antennas are marked as $u_{+}$, $u_{-}$ and $v_{+}$, $v_{-}$, respectively, which are used to determine beam positions in $u$ and $v$ directions. When the beam passes through the BPM chamber, each antenna receives an induced signal. The BPM front-end receiver collects and sends the signal to the regular Hall A DAQ system and another DAQ system designed for parity violation experiments, the HAPPEX system \citep{ADC18bob}. The new BPM receiver was designed by the JLab instrumentation group \citep{mussonece652paper} in order to achieve the required precision at a level of 0.1 mm with a beam current as low as 50 nA. The regular DAQ system was connected to a 13-bit fastbus ADC (Lecroy ADC 1881) with an integration time of 50 ns, which was triggered by a scattered electron event. The HAPPEX system was connected to an 18-bit ADC with an integration time of 875 $\mu$s, which was triggered by a beam helicity signal at 1 kHz. The amplitude, $A$, recorded in the ADC has the following relation with the BPM signal, $\phi$:

\begin{equation}
A\propto\phi\cdot10^{g/20},\label{eq:adcsignalvsantenna}
\end{equation}
where $g$ is the gain of the receiver.

The BPM receiver generates a large time delay for the output signals. The digital filter used in the receiver contributes $1/175\,$s delay time, which was the inverse of the bandwidth setting chosen for the filter. There is a $\sim4\ \mu$s delay as a result of finite processing times. The BPM cannot provide event by event position because of these time delays, due to the 25 kHz fast raster system.

Because of the space limitation between the second chicane magnet and the target, the two BPMs were placed close to each other. One was placed 95.5 cm upstream of the target while the other was placed 69 cm upstream, making the distance between them only 26.5 cm. The short distance magnified the position uncertainty from the BPMs to target.

\subsection{Super harp}

\begin{figure}[tbph]
\begin{centering}
\includegraphics[width=1\columnwidth]{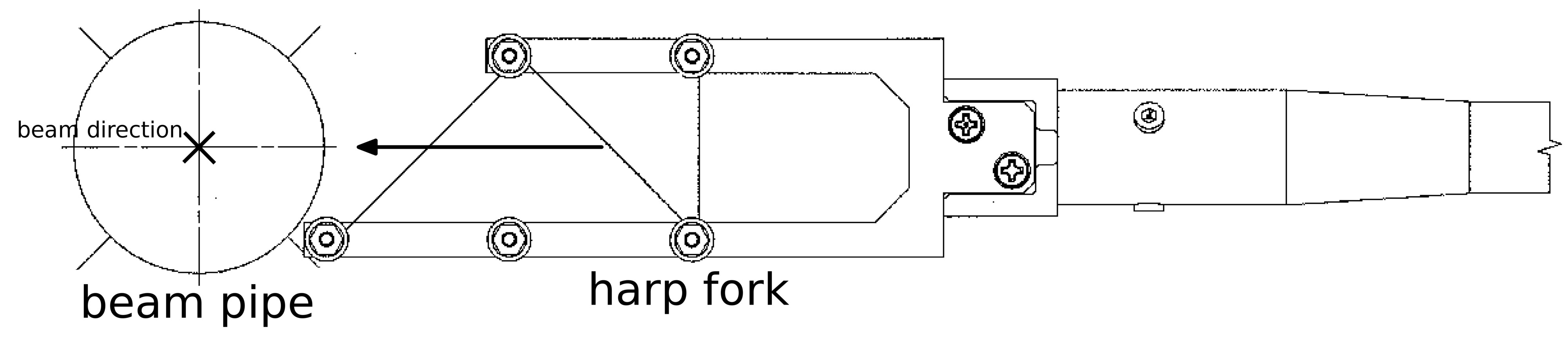} 
\par\end{centering}

\protect\caption{\label{fig:harp-diagram}Harp diagram}
\end{figure}

Two super harps were designed and installed in the beam-line, as shown in Fig.\ref{fig:beamline-for-g2p} (label 6a - 1H04 and 6b - 1H05A), to provide an absolute measurement of the beam position for calibration of the BPMs. The new harps were able to work in pulsed beam (1\% duty factor) with a current of several $\mu$A. A diagram for the harp is shown in Fig.\ref{fig:harp-diagram},
 which consists of three wires with a thickness of 50 $\mu$m, a fork and a controller chassis. The harp chamber is perpendicular to the beam pipe and connected to the beam pipe as part of the vacuum chamber of the beamline. The two harps have different configurations of three wires: vertical($|$), bank left($\backslash$), and bank right($/$) for 1H04, and $/$, $|$, $\backslash$ for 1H05. The angle of the $/$ or $\backslash$ wire is $45^{\circ}$ relative to the wire dock frame. The wires are arranged in a fork (Fig.\ref{fig:harp-diagram}) controlled by a step motor \citep{yan261} which can be moved in and out of the beam-line. The harps must be moved out of the beam-line when production data is being taken because they are invasive to the beam. The original position of the wires was surveyed before the experiment at a precision level of 0.1 mm. As the motor driver moved the fork through the beam, each wire received a signal, which was recorded for further analysis. The signals received from the wire and the step-counters from the motor driver were then sent to an amplifier and the DAQ. The amplification and the speed of the motor were adjustable for the purpose of optimizing the signals of each scan. Recorded data combined with the survey data were used to calculate the absolute beam position.

The signal from the $|$ wire ($peak_{|}$) was used for getting the $x$ position ($x_{harp}$) of the beam, and the signals from the $/$, $\backslash$ wires ($peak_{/}$ and $peak_{\backslash}$) were used for getting the $y$ position ($y_{harp}$): 
\begin{eqnarray}
x_{harp} & = & survey_{|}-peak_{|}\nonumber \\
y_{harp} & = & \frac{1}{2}[(survey_{\backslash}-survey_{/})-(peak_{\backslash}-peak_{/})]\label{eq:harp}
\end{eqnarray}

\subsection{\label{sub:Raster-system}Raster system}

\begin{figure}[tbph]
\begin{centering}
\subfloat[Magnet current waveform for the fast raster in x or y channel]{\protect\begin{centering}
\protect\includegraphics[width=0.4\columnwidth]{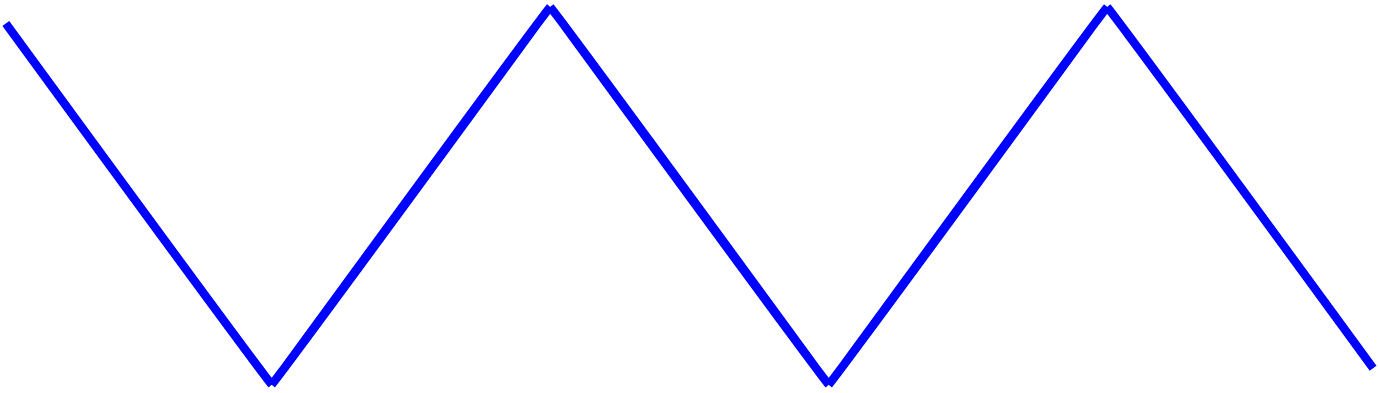}\protect
\par\end{centering}

}$\qquad$\subfloat[The 2D histogram of magnet current signals of fast raster]{\protect\begin{centering}
\protect\includegraphics[width=0.25\columnwidth]{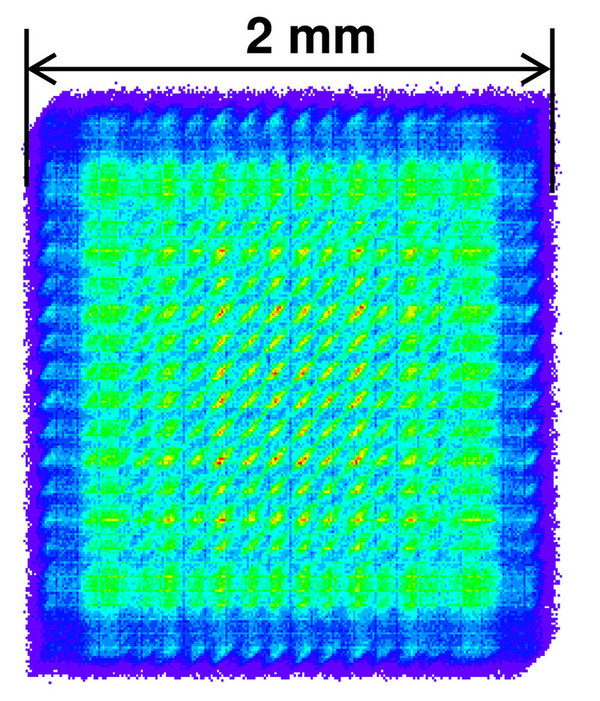}\protect
\par\end{centering}

}
\par\end{centering}

\protect\caption{\label{fig:Fast-raster}Fast raster pattern}
\end{figure}

In order to minimize the depolarization, avoid damage to the target material from radiation, and reduce systematic error for the polarization measurement by NMR (The polarization of the $NH_{3}$ target was measured by using a NMR coil which was placed inside the target cell \citep{Pierce201454}, and the non-uniformity of depolarization could reduce the precision of the NMR measurement due to the measurement being an average over the target), two raster systems were installed at $\sim$17 m upstream of the target, as shown in Fig.\ref{fig:beamline-for-g2p} (labels 2 and 3 for fast and slow rasters, respectively). Both the fast and slow rasters consist of two dipole magnets. The same triangular waveforms with frequency of 25 kHz were used to drive the magnet coils of the fast raster to move the beam in x and y directions, forming a rectangular pattern of about 2 mm$\times$2 mm, as shown in Fig.\ref{fig:Fast-raster}. 

A dual-channel function-generator\footnote{agilent 33522A function generator, http://www.home.agilent.com/en/pd-1871286-pn-33522A/function-arbitrary-waveform-generator-30-mhz} was used to generate two independent waveforms to drive the magnet coils of the slow raster. The waveforms for the $x$ and $y$ directions are:

\begin{eqnarray}
x & = & A_{x}t{}^{1/2}sin(\omega t),\nonumber \\
y & = & A_{y}(t+t_{0})^{1/2}sin(\omega t+\mbox{\ensuremath{\phi}}),\label{eq:slraster_func}
\end{eqnarray}
where the $A_{x}$ and $A_{y}$ are the maximum amplitude, $t_{0}$ and $\phi$ are the AM and sin phase difference between $x$ and $y$ waveform, respectively. 
\begin{figure}[tbph]
\begin{centering}
\subfloat[Magnet current waveform for the slow raster]{\protect\begin{centering}
\protect\includegraphics[width=0.7\columnwidth]{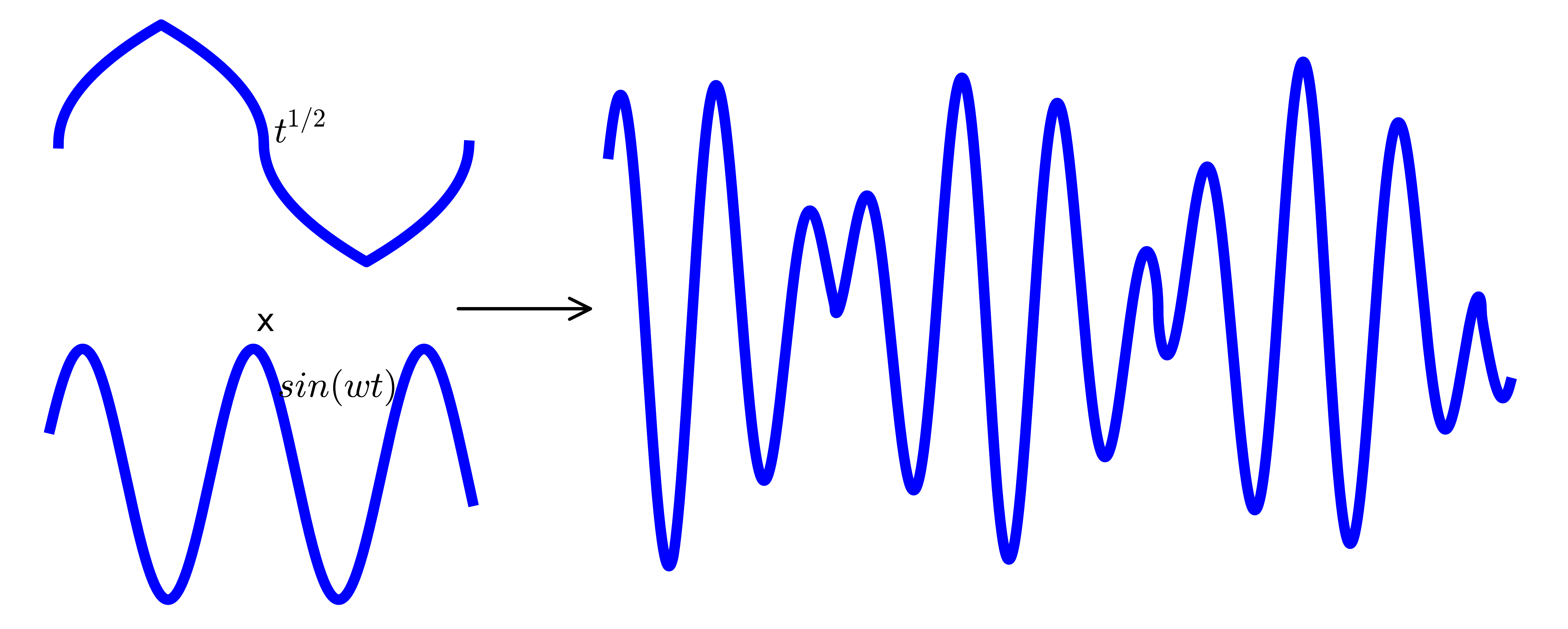}\protect
\par\end{centering}

}\subfloat[The 2D histogram of magnet current signals of slow raster]{\protect\begin{centering}
\protect\includegraphics[width=0.3\columnwidth]{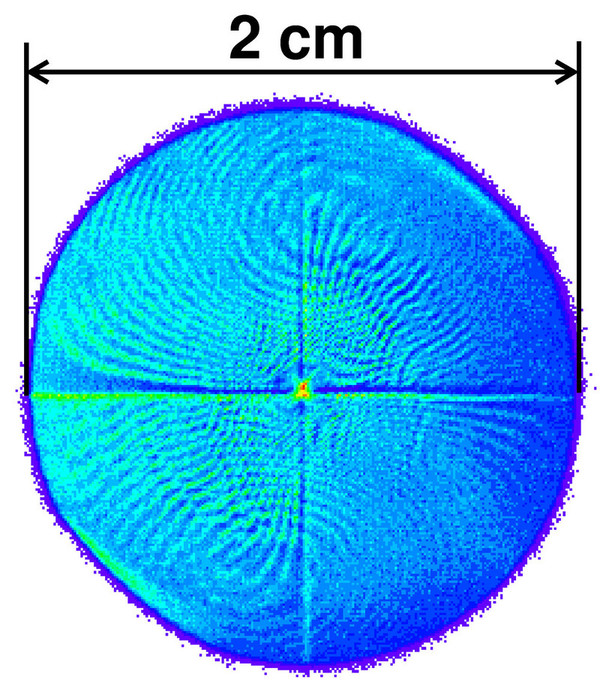}\protect
\par\end{centering}

}
\par\end{centering}

\protect\caption{\label{fig:Slow-raster}Slow raster pattern }
\end{figure}
 Both of them are sine functions modulated by a function $t^{1/2}$ in order to generate a uniform circular pattern \citep{HCraster200501}, as shown in Fig.\ref{fig:Slow-raster}. The frequencies of the $x$ and $y$ waveforms kept same: $\omega=99.412$ Hz. In order to cycle the amplitude modulation (AM) function, four piece-wise functions are combined together. The first term is $t^{1/2}$, and the second term is $period-t^{1/2}$, and so on for the third and fourth terms. The cycled function has the frequency of 30 Hz. 
\begin{figure}[tbph]
\begin{centering}
\subfloat[\label{fig:AM-phase-=00003D0022600}Raster pattern with $t_{0}$\ensuremath{\neq}0]{\protect\begin{centering}
\protect\includegraphics[width=0.33\columnwidth]{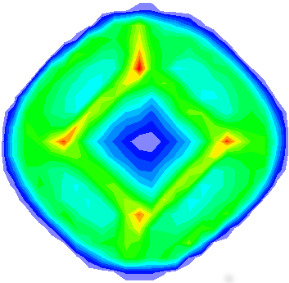}\protect
\par\end{centering}

}\subfloat[\label{fig:simulated-raster-shape}Simulated raster pattern, with $t_{0}$\ensuremath{\neq}0]{\protect\begin{centering}
\protect\includegraphics[width=0.33\columnwidth]{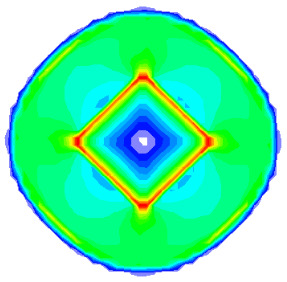}\protect
\par\end{centering}

}\subfloat[\label{fig:AM-phase=00003D00003D0}Manually adjust $t_{0}$ to 0]{\protect\begin{centering}
\protect\includegraphics[width=0.33\columnwidth]{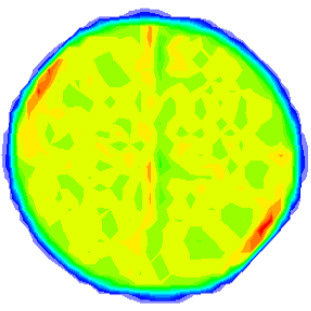}\protect
\par\end{centering}

}
\par\end{centering}

\protect\caption{\label{fig:Slow-raster-uniformity}None-zero $t_{0}$ caused slow raster non-uniformity, (a) and (c) are from the data recorded in the ADC, (b) is simulated. The color palette shows the uniformity of the raster pattern.}
\end{figure}

The $\phi$ was locked to $\frac{\pi}{2}$ by the function generator, while the $t_{0}$ was manually fixed to 0. Non-zero $t_{0}$ could cause a non-uniformity pattern, as shown in Fig.\ref{fig:Slow-raster-uniformity}(a), which would cause non-uniformity beam distribution. A simulation was reproduced the non-uniformity by setting the $t_{0}$ to non-zero, as shown in Fig.\ref{fig:Slow-raster-uniformity}(b). The $t_{0}$ was carefully adjusted and minimized before production data taking to avoid the non-uniformity. The pattern of the spread beam was relatively uniform after this adjustment during the experiment, as shown in Fig.\ref{fig:Slow-raster-uniformity}(c).

\section{Data analysis}

\subsection{Harp scans for measuring absolute beam position}

\begin{figure}[tbph]
\begin{centering}
\subfloat[\label{fig:position-vs-index}position vs index for harp scan, used for extending position record]{\protect\begin{centering}
\protect\includegraphics[width=0.4\columnwidth]{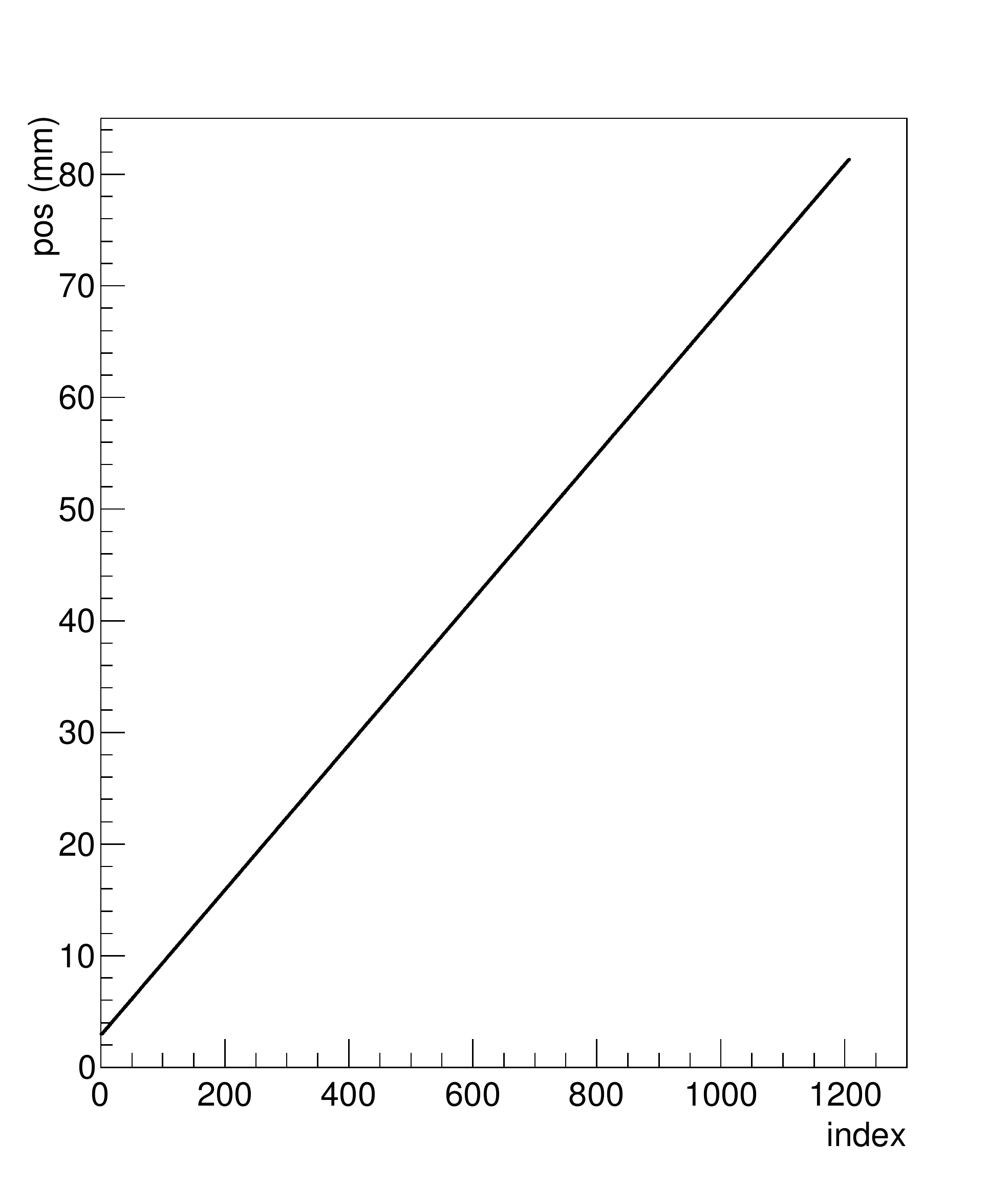}\protect
\par\end{centering}

}$\qquad$\subfloat[\label{fig:position-vs-signal}Signal vs position for harp, $x$ axis is position, $y$ axis is the strength of signal, which is the ADC value. ]{\protect\begin{centering}
\protect\includegraphics[width=0.4\columnwidth]{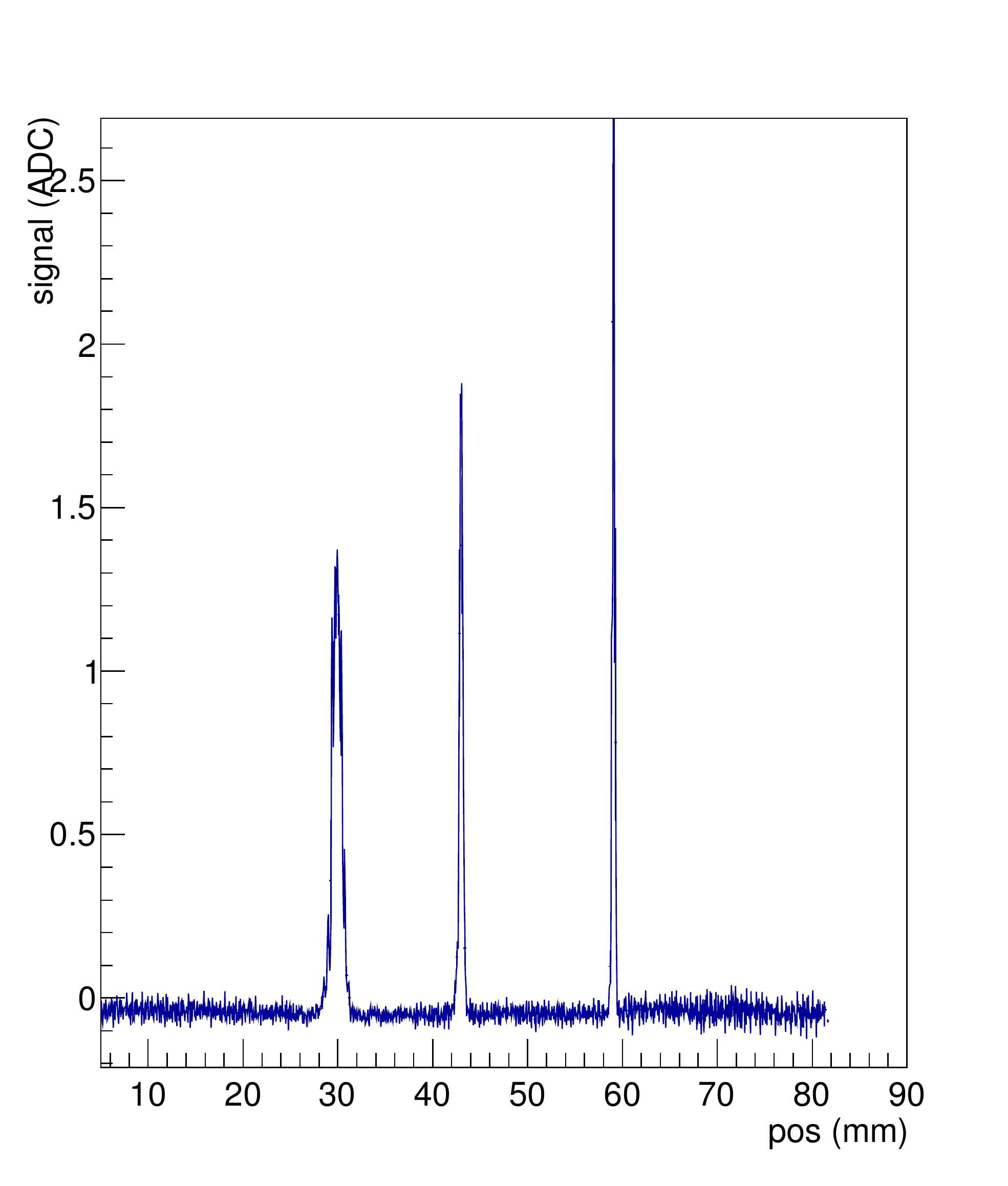}\protect
\par\end{centering}

}
\par\end{centering}

\protect\caption{\label{fig:1H05A-harp-scan}1H05A harp scan data}
\end{figure}

An example of a harp scan result is shown in Fig.\ref{fig:1H05A-harp-scan}.
 There are three groups of recorded data for each harp scan, which are ``index'', ``position'', and ``signal''. The index is related to the moving steps of the fork during the scan. Each step of the index increases by 0.008-0.07 mm depending on the speed of the motor driver \citep{yan261}. The position is the wire location for each index. The testing results show a good linear relation between the position and the index as shown in Fig.\ref{fig:1H05A-harp-scan}(a), because the motor speed is uniform. The line is the fitted result with $pos=a*index+b$. According to this linear relation, interpolation or extrapolation can be applied when a few data points are missing, in some cases. The strength of signal vs. position is plotted in Fig.\ref{fig:1H05A-harp-scan}(b). Each peak represents the location when one of the three wires passed through the beam.

The positions measured by the two harps were used for calibrating the beam positions in the two BPMs. When the chicane magnets were on, beam did not pass straight through from the first harp to the second harp. BPM calibrations using two harps were only possible when the chicane magnets were off, i.e. in the straight-through settings. Since the BPM was calibrated in the local coordinate system, the calibration constants were independent from the settings of other instruments. To make sure that the calibration constants for the BPMs were still valid during the non-straight-through settings, the settings for the BPM receiver were kept the same as in the straight-through settings during production running.

The scan data from the harps were not reliable when the current of CW beam (100\% duty factor) was lower than 100 nA due to the low signal-to-noise ratio. The harp scans were taken in pulsed mode at a current of a few $\mu$A, while the BPMs were used for production data taking in CW mode at a beam current of 50-100 nA. For a BPM calibration run, a harp scan was done first in pulsed mode, then a DAQ run was taken immediately to record the ADC value in CW mode without changing the beam position. The harp scan was then taken again in the pulsed mode to double check the beam position. The harp scan data was discarded and the scan was taken again if the beam position changed.

\subsection{BPM data analysis and calibration}

\begin{figure}[tbph]
\begin{centering}
\subfloat[Mirror method]{\protect\begin{centering}
\protect\includegraphics[width=0.4\columnwidth]{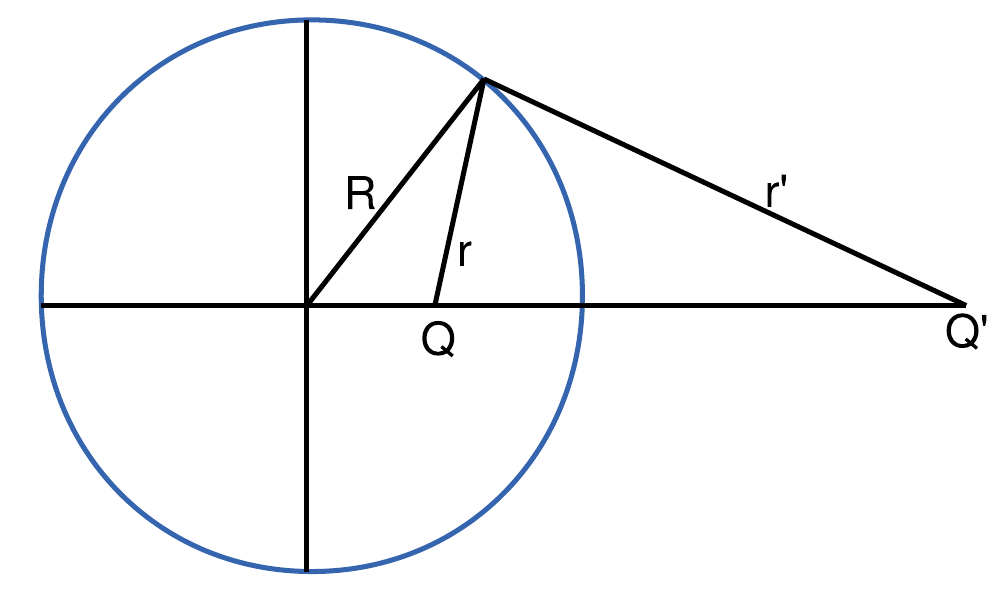}\protect
\par\end{centering}

}\subfloat[Signal on antenna]{\protect\begin{centering}
\protect\includegraphics[width=0.4\columnwidth]{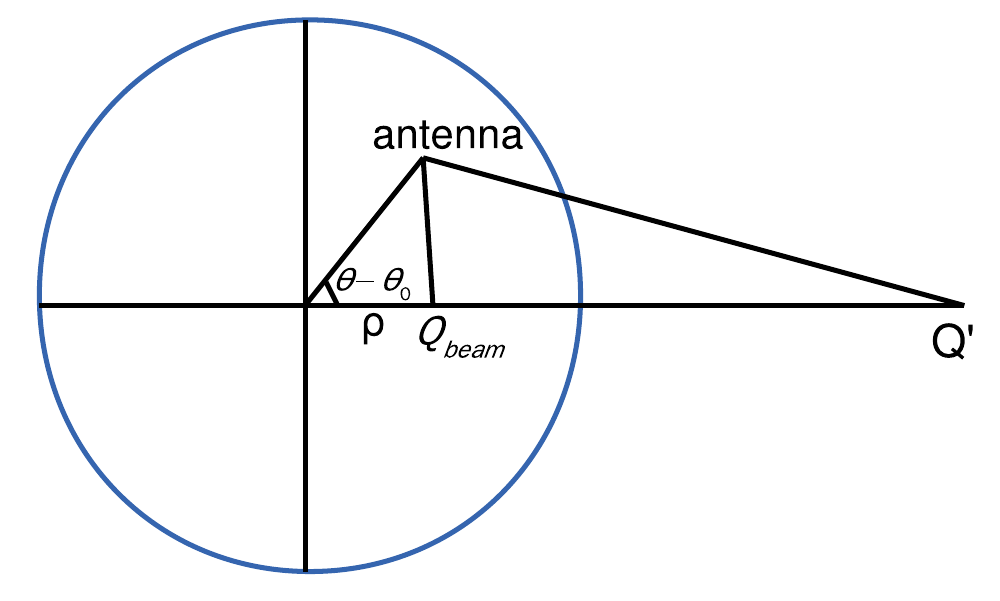}\protect
\par\end{centering}

}
\par\end{centering}

\protect\caption{\label{fig:mirror_method}Signal for each antenna of BPM}
\end{figure}

The traditional difference-over-sum ($\Delta/\Sigma$) method of calculating the beam position has the non-linearity effect at the position far away from the center of the beam pipe \citep{Barry301}. It is necessary to correct the equation of $\Delta/\Sigma$ since we have a slow raster with a large size of \textasciitilde{}2 cm. With the assumption of an infinitely long chamber and neglecting the antenna influence on the electric field inside the chamber, the signal from each antenna excited by the beam can be calculated via image charge method (Fig.\ref{fig:mirror_method}) \citep{carmanbpm,piotbpm} : 

\begin{equation}
\phi_{i}=\phi_{0}I\frac{R^{2}-\rho^{2}}{R^{2}+\rho^{2}-2R\rho cos(\theta_{i}-\theta_{0})},\label{eq:antenna_signal}
\end{equation}
where $\phi_{i}$ is the signal received in the antenna, and $i$ is $u_{+}$, $u_{-}$, $v_{+}$ and $v_{-}$, respectively, $\phi_{0}$ is a constant related to the geometry of the BPM-chamber and the output resistance, $I$ is the beam current, $R$ is the radius of the BPM vacuum chamber, $\rho$ is the radial position of the beam, and $\theta_{i}-\theta_{0}$ is the angle difference between the antenna and the beam in the polar coordinate . 

In order to extract the beam position information, and eliminate the dependence on the beam current in equation (\ref{eq:antenna_signal}), the $\Delta/\Sigma$ method is used as follows:
\begin{eqnarray}
D_{U} & = & \frac{\phi_{U+}-\phi_{U-}}{\phi_{U+}+\phi_{U-}},\label{eq:diff/sum}
\end{eqnarray}
where $U$ denotes $u$ and $v$. Substituting equation (\ref{eq:antenna_signal}) into equation (\ref{eq:diff/sum}), it can be rewritten as follows:

\begin{equation}
D_{U}=\frac{\phi_{U+}-\phi_{U-}}{\phi_{U+}+\phi_{U-}}=\frac{2}{R}\frac{\rho cos(\theta-\theta_{0})}{1+\frac{\rho^{2}}{R^{2}}}=\frac{2}{R}\frac{U}{1+\frac{\rho^{2}}{R^{2}}},\label{eq:xb=00003Dxy}
\end{equation}
where $\rho^{2}=u^{2}+v^{2}$. When $u^{2}+v^{2}\ll R^{2}$, equation (\ref{eq:xb=00003Dxy}) is simplified as:
\begin{eqnarray}
U & \approx & \frac{R}{2}D_{U}=\frac{R}{2}\frac{\phi_{U+}-\phi_{U-}}{\phi_{U+}+\phi_{U-}}.\label{eq:litexyb=00003Dxy}
\end{eqnarray}
Equation (\ref{eq:litexyb=00003Dxy}) can be used in the simple case when the beam is near the center of the beam pipe. When the beam is far from the center, equation (\ref{eq:litexyb=00003Dxy}) is no longer valid. For the g2p experiment, the beam was rastered to have a diameter of about 2 cm at the target. From equation (\ref{eq:xb=00003Dxy}) the beam position is calculated as:
\begin{eqnarray}
U & = & RD_{U}(\frac{1}{D_{u}^{2}+D_{v}^{2}}-\frac{1}{\sqrt{D_{u}^{2}+D_{v}^{2}}}\sqrt{\frac{1}{D_{u}^{2}+D_{v}^{2}}-1}).\label{eq:realxy}
\end{eqnarray}

\begin{figure}
\begin{centering}
\subfloat[Data from experiment]{\protect\begin{centering}
\protect\includegraphics[width=0.45\columnwidth]{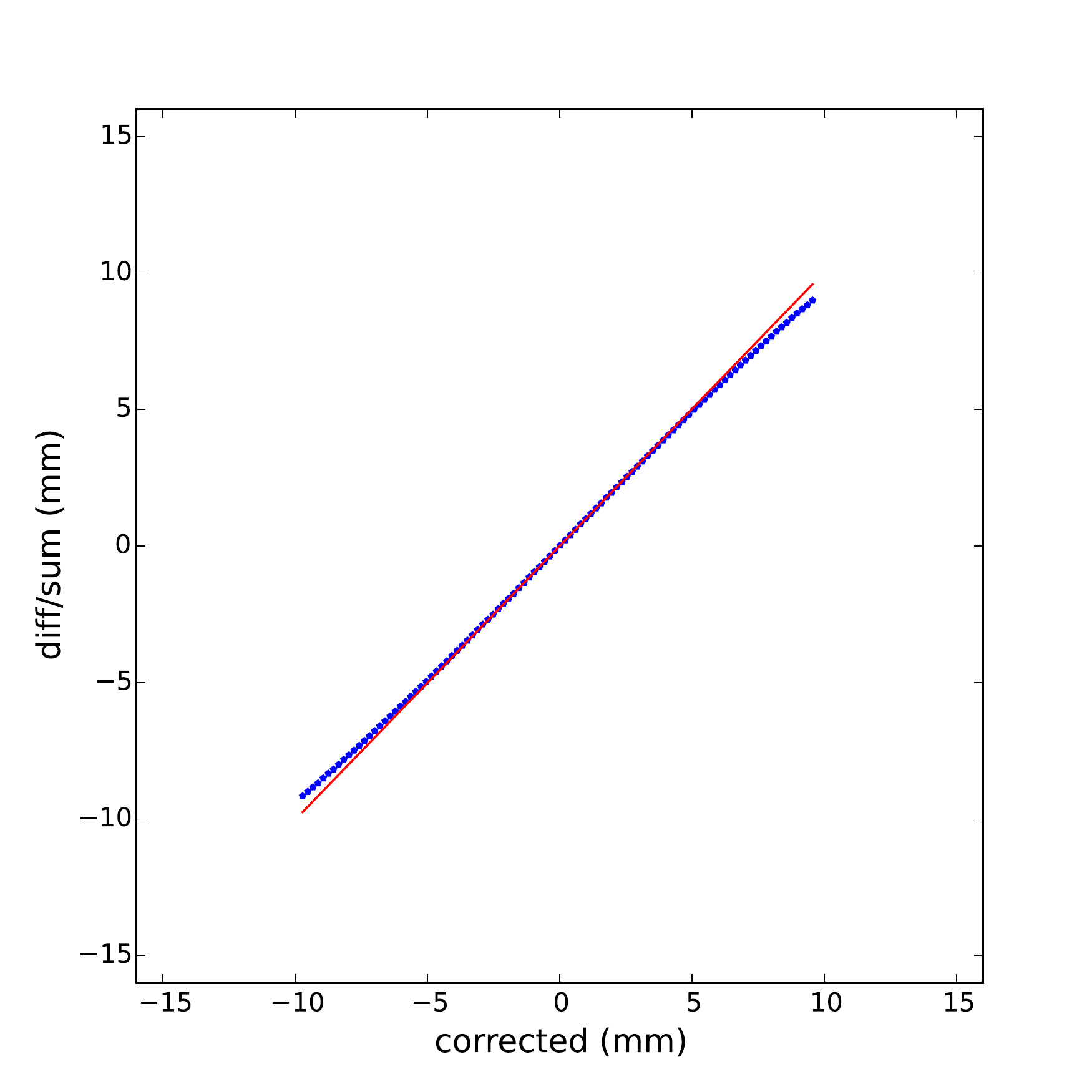}\protect
\par\end{centering}

}\subfloat[Data from bench test]{\protect\begin{centering}
\protect\includegraphics[width=0.45\columnwidth]{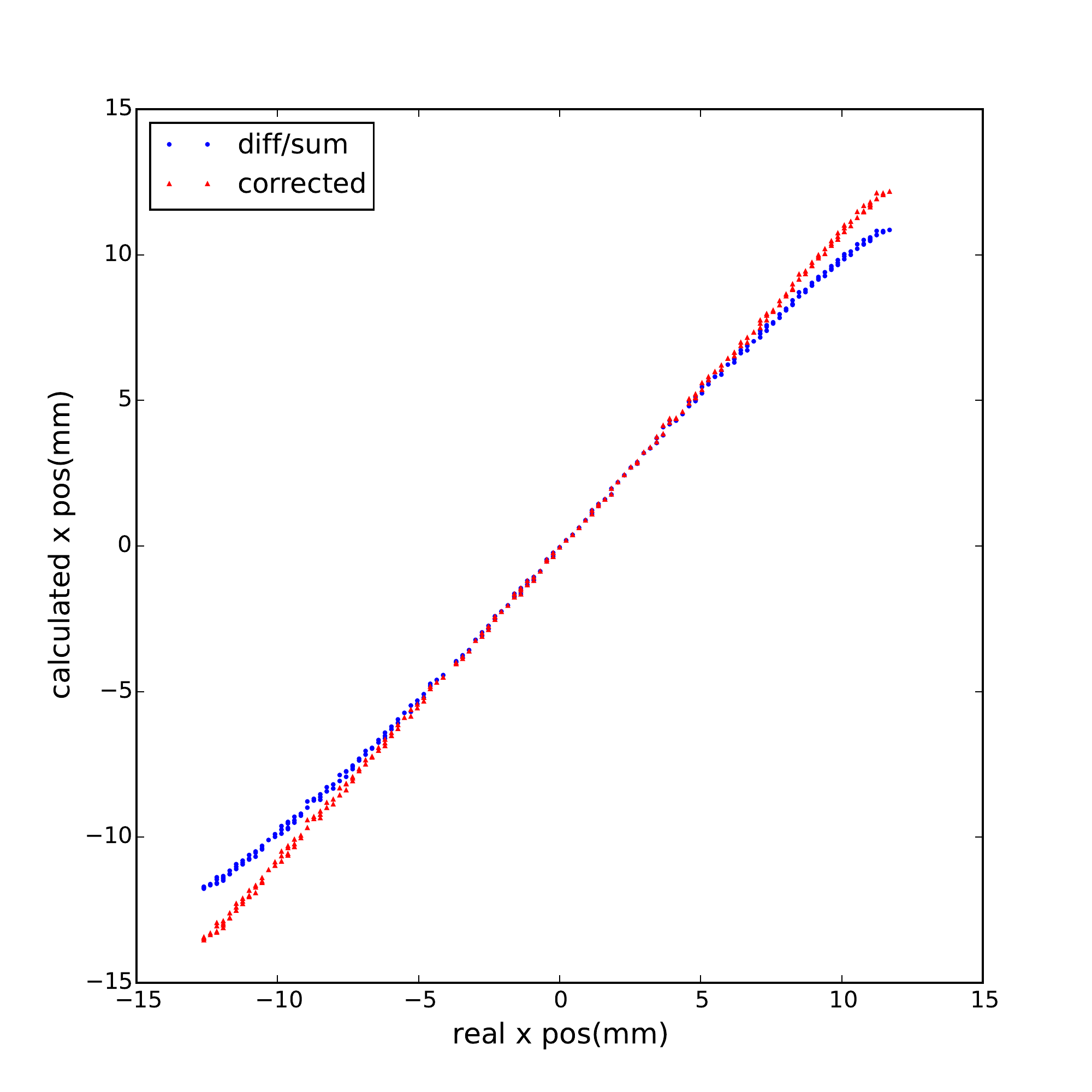}\protect
\par\end{centering}

}
\par\end{centering}

\protect\caption{\label{fig:BPM-non-linearity-correction}BPM non-linearity correction. (a) Comparison between the position calculated from $\Delta/\Sigma$ equation (\ref{eq:litexyb=00003Dxy}) (y axis) and the one from correction equation (\ref{eq:realxy}) (x axis). Red solid line is a reference line came from linear fit of the center points. Data is collected from the experiment. (b) Comparison between the $\Delta/\Sigma$ equation (\ref{eq:litexyb=00003Dxy}) and the correction equation (\ref{eq:realxy}) using the bench test data. The x axis is the real beam position. The red triangles are the positions calculated from correction equation (\ref{eq:realxy}). The blue circles are the positions calculated from $\Delta/\Sigma$ equation (\ref{eq:litexyb=00003Dxy}) .}
\end{figure}
The correction equation is tested by using the experiment data and the bench test data. Fig.\ref{fig:BPM-non-linearity-correction}(a) shows the comparison between the position calculated from the correction equation (\ref{eq:realxy}) and the one from the $\Delta/\Sigma$ equation (\ref{eq:litexyb=00003Dxy}). The red solid line is a reference line came from linear fit of the center points. Fig.\ref{fig:BPM-non-linearity-correction}(b) shows the comparison with the real beam position from the bench test data. In this way the method using equation (\ref{eq:realxy}) can correct the non-linearity effect caused by the $\Delta/\Sigma$ method. The handling of the BPM information which only used for the center beam position (discussed in chapter \ref{sub:Det-ebe-pos}) also reduced this non-linearity effect. 

The final information recorded in the data-stream was designed to have a linear response with the raw signal in the 50-100nA current range. The $\phi_{i}$ in equation (\ref{eq:xb=00003Dxy}) can be rewritten as:

\begin{equation}
\phi_{i}=a_{i}(A_{i}-A_{i\_ped}+b_{i}),\label{eq:phivsA}
\end{equation}
where $A_{i}$ and $A_{i\_ped}$ are the recorded ADC value and pedestal value, and $a_{i}$ and $b_{i}$ are the slope and intercept of the relationship between $\phi_{i}$ and $A_{i}-A_{i\_ped}$. Equation (\ref{eq:litexyb=00003Dxy}) can be rewritten as:

\begin{eqnarray}
D_{U} & = & \frac{(A_{U+}-A_{U+\_ped}+b_{U+})-h_{U}(A_{U-}-A_{U-\_ped}+b_{U-})}{(A_{U+}-A_{U+\_ped}+b_{U+})+h_{U}(A_{U-}-A_{U-\_ped}+b_{U-})},\label{eq:xbfab}
\end{eqnarray}
where $h_{U}$ = $a_{U-}/a_{U+}$, which is related to the ratio of the signals from the $U_{+}$ and $U_{-}$ antennas and the gain settings of the two channels. 

Combining the equations (\ref{eq:phivsA}) and (\ref{eq:antenna_signal}), the calibration constant $b_{i}$ was obtained by taking the linear fit between the ADC values of BPM and the beam current: $I\propto(A_{i}-A_{i\_ped}+b_{i})$. Besides, the linear fit used a group of runs which had the same beam position but different beam current.
\begin{figure}[tbph]
\begin{centering}
\includegraphics[width=0.5\columnwidth]{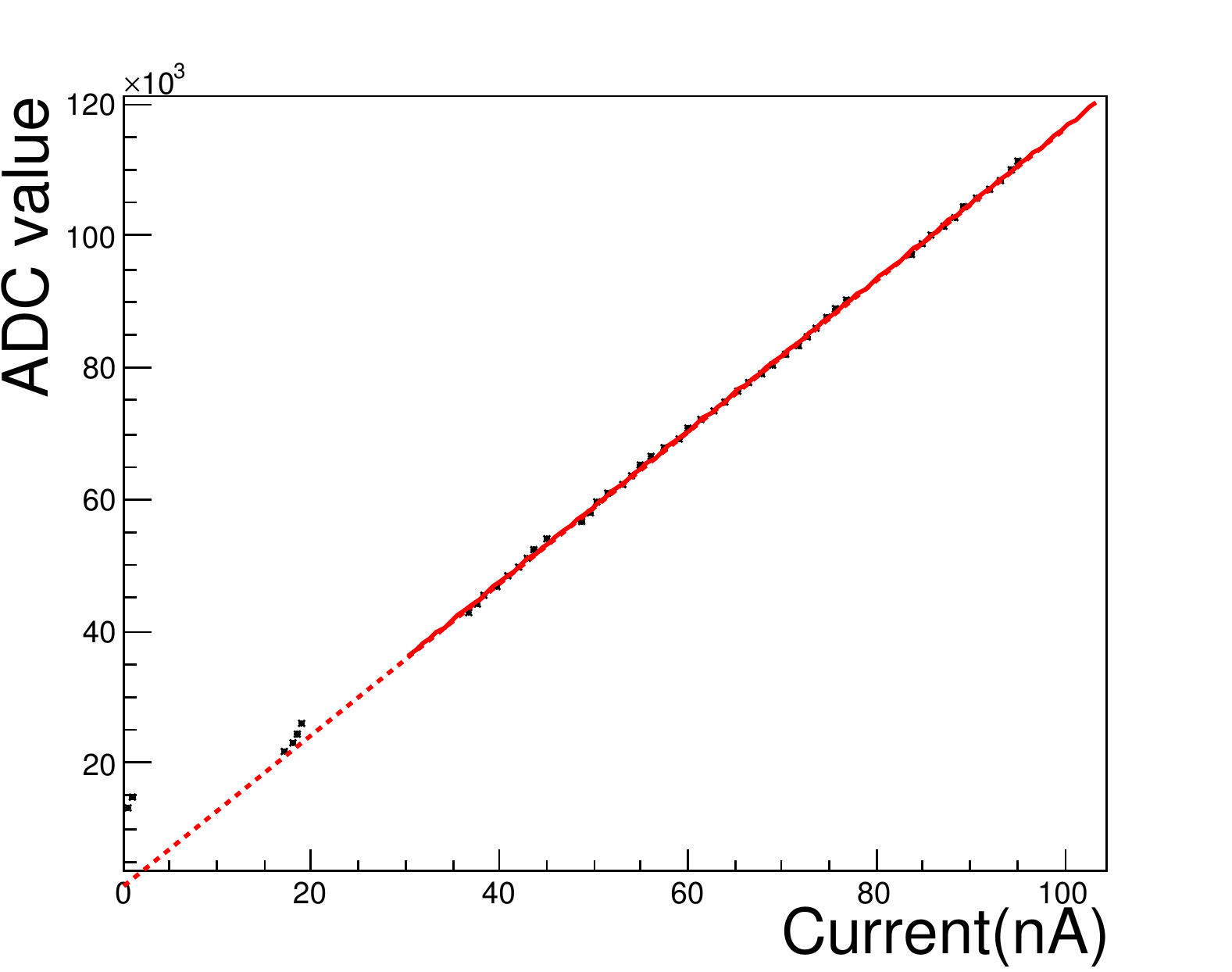} 
\par\end{centering}

\protect\caption{\label{fig:BPM-raw-signalVScurrent}ADC value of BPM raw signal ($A-A_{ped}$) V.S. beam current}
\end{figure}
 Figure \ref{fig:BPM-raw-signalVScurrent} shows the $A_{i}-A_{i\_ped}$ versus the beam current. It shows that the ADC values were linear with beam current in the considering current range of 50-100 nA. The intercept from the linear fit of Fig.\ref{fig:BPM-raw-signalVScurrent} is the value $b_{i}$. 

\begin{figure}[tbph]
\begin{centering}
\includegraphics[width=0.4\columnwidth]{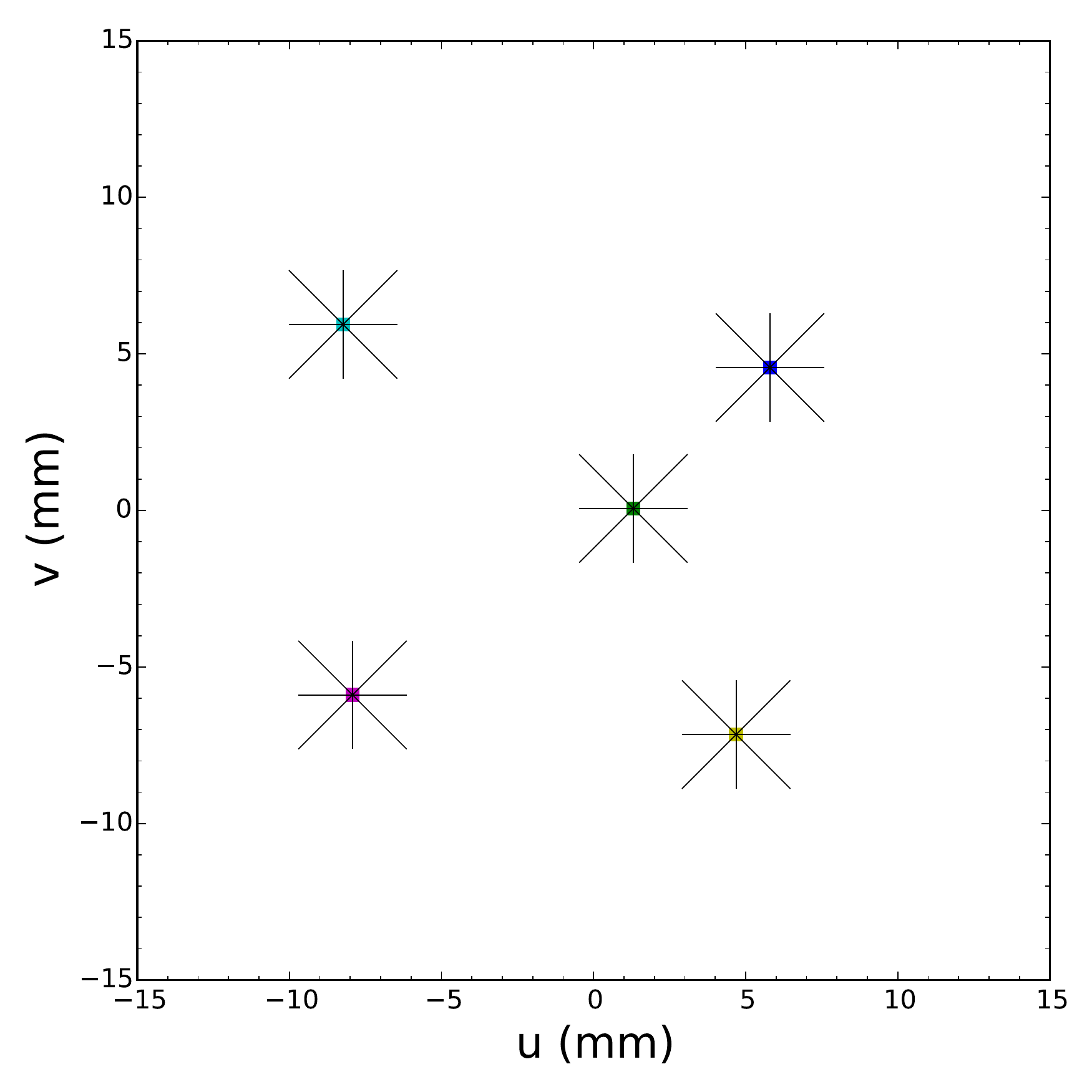}
\par\end{centering}

\protect\caption{\label{fig:Harpscan}Harp scan data combined with BPM data, the asterisks are the positions from harp, while the dots are from BPM. }
\end{figure}

By transporting the position $x_{harp}$ and $y_{harp}$ in equation (\ref{eq:harp}) from two harps to the BPM local coordinate $u_{harp}$ and $v_{harp}$, a fit between the BPM data $U$ and the harp data $U_{harp}$ determined three calibration constants $c_{0}$, $c_{1}$ and $c_{2}$:

\begin{equation}
U_{harp}=U_{c}=c_{0}+c_{1}u+c_{2}v,\label{eq:Uvu}
\end{equation}
where $U_{c}$ is the calibrated BPM position. It was converted to Hall coordinate $X_{c}$ for further transporting to the target location. An calibration example is shown in Fig.\ref{fig:Harpscan}. 
 The asterisks and the dots in Fig.\ref{fig:Harpscan} represent $U_{harp}$ and $U$, respectively. 

In order to reduce the noise and improve the resolution during data analysis, a software filter was applied. Since the 18 bit ADC was triggered by the helicity signal with a fixed frequency, it could be regarded as a sampling ADC. 
\begin{figure}[tbph]
\begin{centering}
\subfloat[Normal run with beam]{\protect\begin{centering}
\protect\includegraphics[width=0.95\columnwidth]{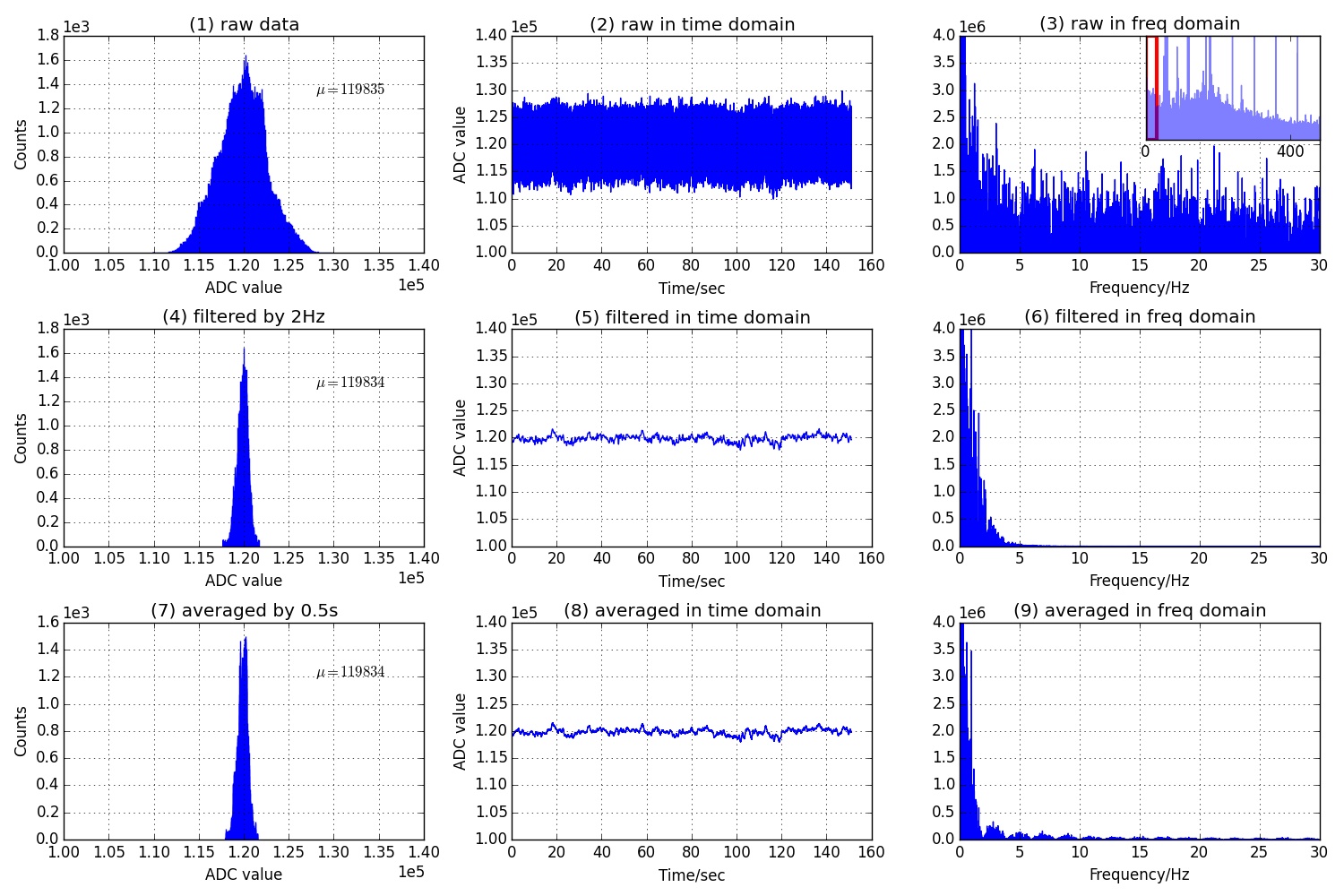}\protect
\par\end{centering}

}
\par\end{centering}

\begin{centering}
\subfloat[Pedestal run without beam]{\protect\begin{centering}
\protect\includegraphics[width=0.95\columnwidth]{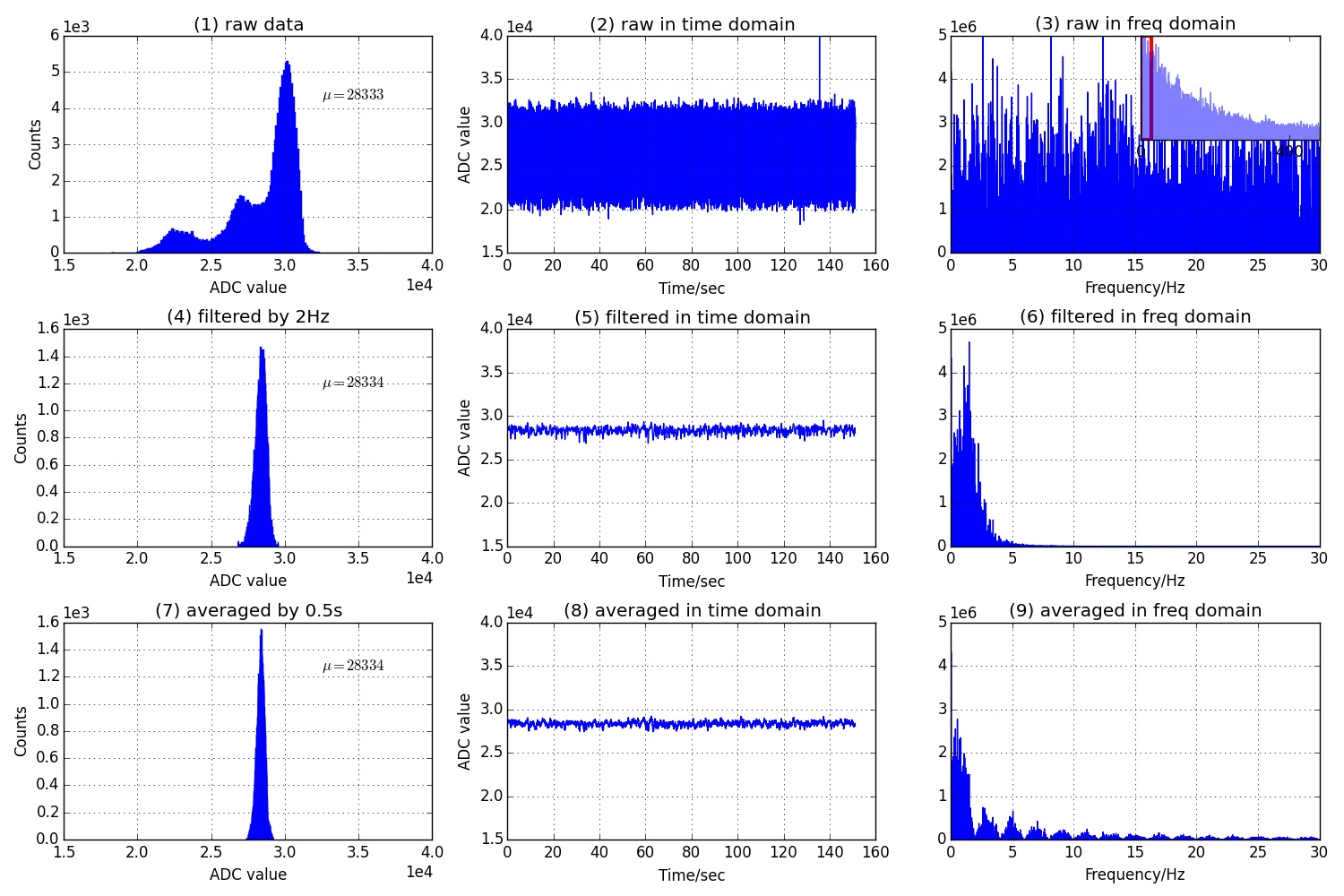}\protect
\par\end{centering}

}
\par\end{centering}

\protect\caption{\label{fig:2Hz-filter-added}Software filter applied to BPM signal. (a) is the signal with beam, (b) is the pedestal signal without beam. (1,2,3) in (a,b) are the raw signal without applying the filter, (4,5,6) are applied a 2 Hz finite-impulse-response filter with 4th order. (7,8,9) are averaged with 0.5 s. (1,4,7) are the 1-D histogram of the recorded signal, (2,5,8) are the signal in time domain, (3,6,9) are in frequency domain. Note all of the plots in (a) are from a single signal, same as in (b).}
\end{figure}
 Fig.\ref{fig:2Hz-filter-added} shows the signal dealt with a 2 Hz low pass filter. Three plots at the bottom of Fig.\ref{fig:2Hz-filter-added} (a,b) are the averaged signal used for comparing with the filtered signal. The results show that the 2 Hz filter and the 0.5 s average are consistent within the required precision. The filter also erases the beam displacement caused by the rasters, which is necessary to extract the position of the beam center.

\subsection{Beam position reconstruction at the target\label{sub:Beam-position-reconstruction}}

\begin{figure}[tbph]
\begin{centering}
\includegraphics[width=0.4\columnwidth]{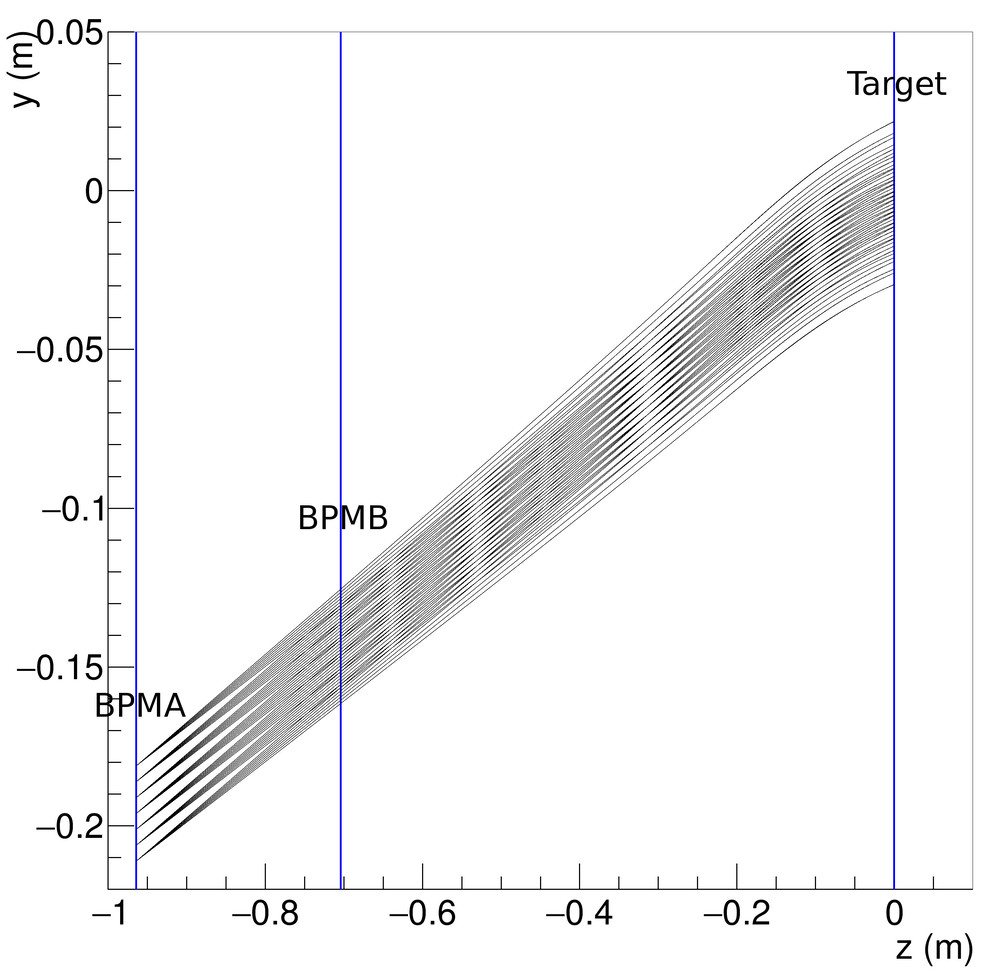} 
\par\end{centering}

\protect\caption{\label{fig:Transporting-beam-position}Transporting beam position from BPM to target with transverse target magnet field. Trajectories are from simulation. Blue lines show the z positions of BPMA, BPMB and target. y and z are in global Hall coordinate.}
\end{figure}

It is easy to transport the position from the BPMs to the target by using a linear transportation method for the straight through setting. For the settings with a transverse magnetic field at the target, the linear transportation method cannot be used since the beam is bent near the target. A target magnet field map \citep{winestosca} was generated from the TOSCA model. To test the accuracy of the TOSCA model, the target magnet field was measured before the experiment \citep{jiefieldmap,chaofieldmap}. The position and angle at target were calculated in terms of the positions at BPMA and BPMB, together with the magnet field information. Fits were used to speed up the calculation. To do this, a full simulation was taken by generating thousands of trajectories with different initial positions and angles. The fits were compared with the full simulation and they are consistent with negligible difference. Fig.\ref{fig:Transporting-beam-position} shows the trajectories from the simulation. 

The fitted transport functions were only used to transport the beam center position from the two BPMs to the target by applying the 2 Hz filter, which filtered out the fast raster and slow raster motion to keep only the beam center position. The transported position were expressed as $X_{center}$.

\subsection{\label{sub:Det-ebe-pos}Determining the beam position event-by-event}

\begin{figure}
\begin{centering}
\includegraphics[width=0.6\columnwidth]{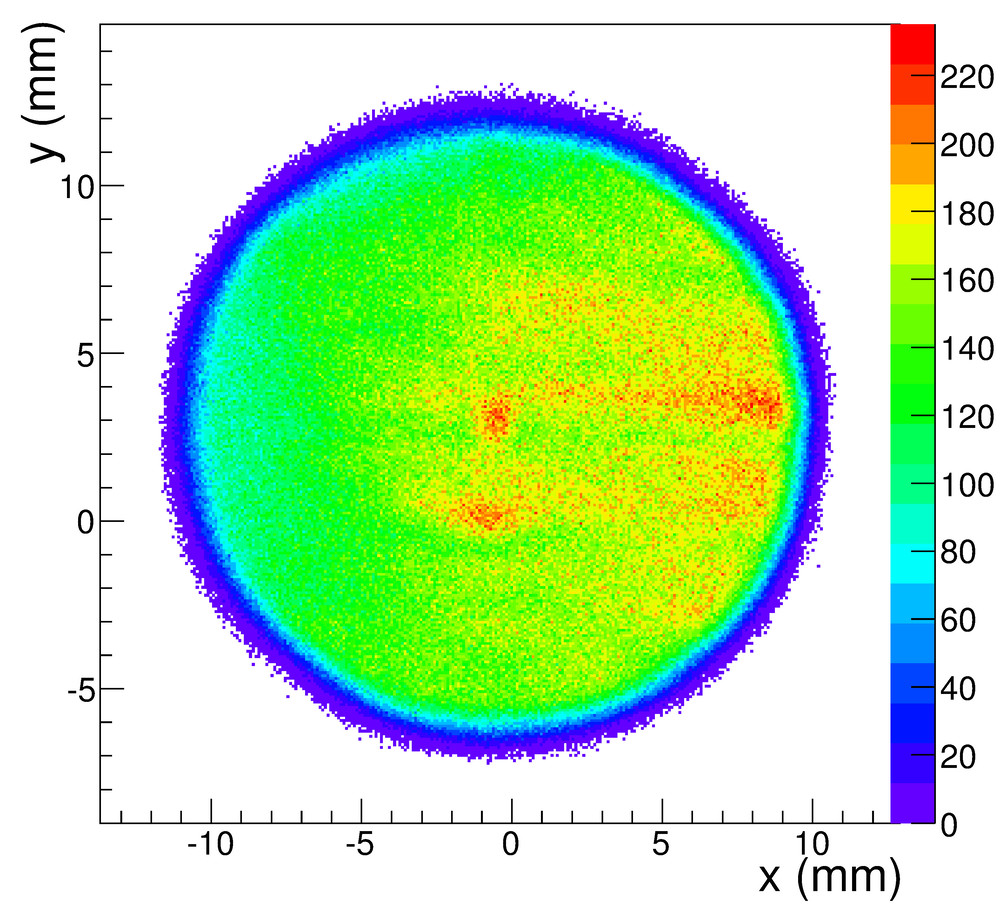}
\par\end{centering}

\protect\caption{\label{fig:Reconstructed-beam-position}Reconstructed beam position at the target}

\end{figure}

The readout of the magnet current for the two rasters was connected to a series of ADCs. Two scintillator planes in the HRS form a DAQ trigger. This pulse signal triggered the ADC to record the raster magnet current for each event. The information from the rasters and the BPMs was combined to provide the beam position event-by-event. The position at the target was determined as:

\begin{eqnarray}
X & = & X_{center}+X_{fstraster}+X_{slraster},\label{eq:tot_pos}
\end{eqnarray}
where $X_{fstraster}$ and $X_{slraster}$ were the position displaced by the fast raster and slow raster, respectively, which were converted from the current values of the two raster magnets. The calibration of the conversion factors between the magnet current of the rasters and the displaced position will be discussed in the next subsection. An example of reconstructed beam position is shown in Fig.\ref{fig:Reconstructed-beam-position}.

\subsubsection{Conversion factor for the slow raster}

\begin{figure}[tbph]
\begin{centering}
\subfloat[\label{fig:slow-raster-run_different_size}ADC value of slow raster, with the raster current changing]{\protect\begin{centering}
\protect\includegraphics[width=0.33\columnwidth]{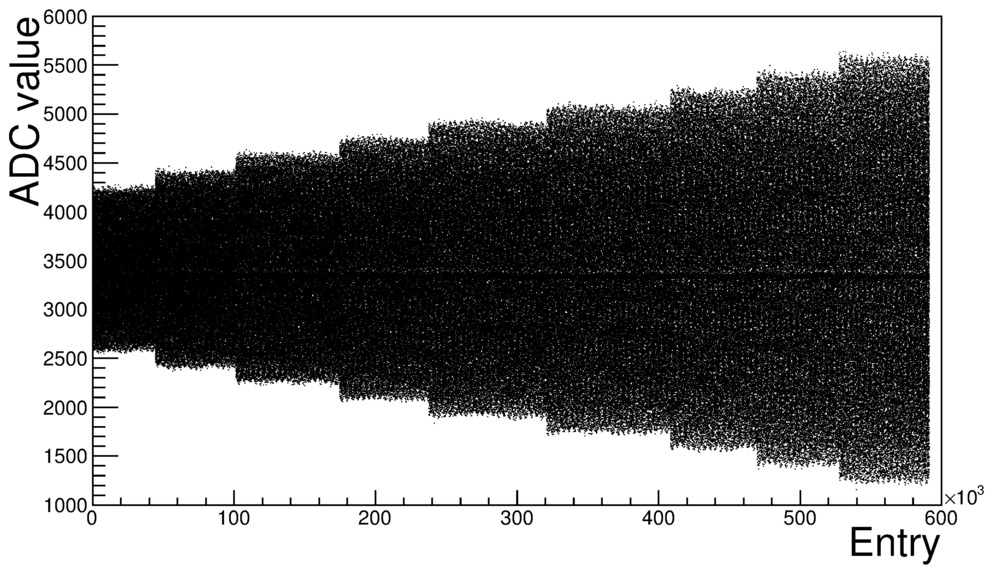}\protect
\par\end{centering}

}$\quad$\subfloat[\label{fig:slow-raster-oval-fit}Elliptical fit for the spread of magnet current of slow raster]{\protect\begin{centering}
\protect\includegraphics[width=0.28\columnwidth]{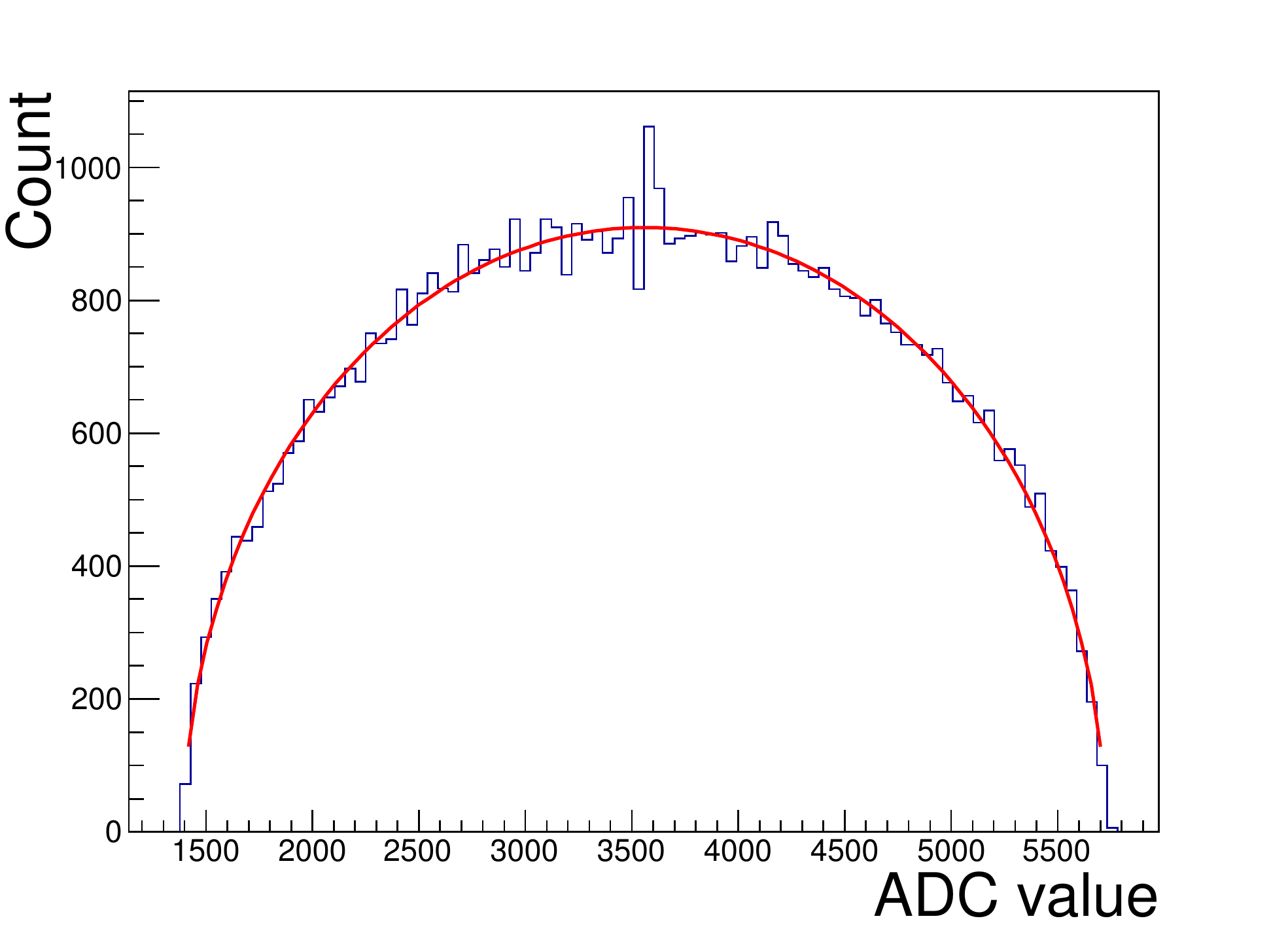}\protect
\par\end{centering}

}$\quad$\subfloat[\label{fig:slow-raster-size_b}Linear fit between the raster current and the range of beam distribution]{\protect\begin{centering}
\protect\includegraphics[width=0.3\columnwidth]{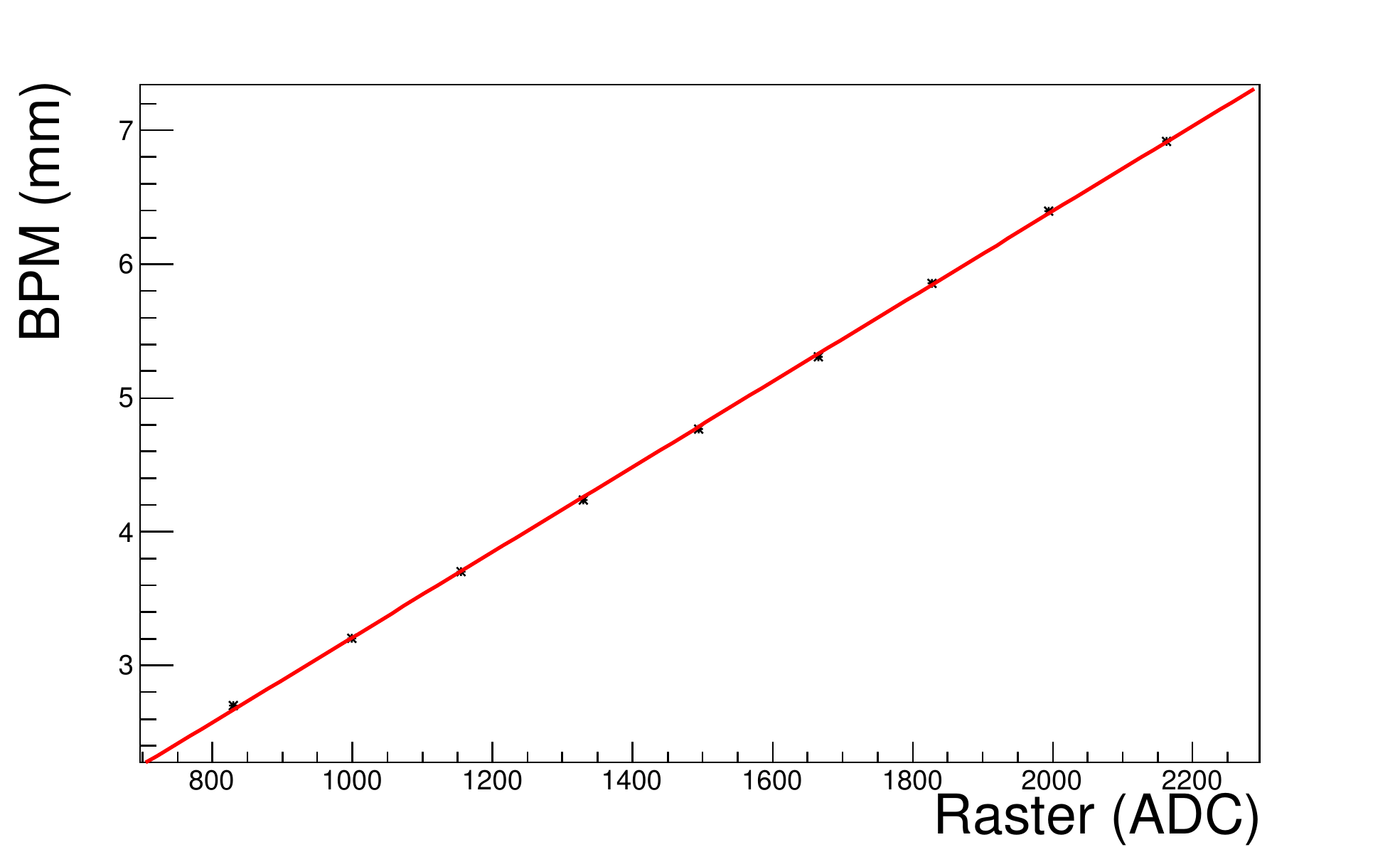}\protect
\par\end{centering}

}
\par\end{centering}

\protect\caption{\label{fig:Slow-raster-size}Converting the raster current to beam position shift}
\end{figure}

Two methods were used to calibrate the conversion factor for the slow raster. The first method used the calibrated BPM information, i.e., comparing the raster magnet current with the beam shape shown in the ADC of the BPMs. Several calibrations were taken during different run periods at a beam current of 100nA using different values of the raster magnet current, as shown in Fig.\ref{fig:Slow-raster-size}(a). 

The range of the beam distribution at the target was calculated from the ranges at the two BPMs without applying the filter, using the transport functions fitted previously. The range of the beam distribution at the two BPMs and the amplitude of the raster current were calculated from elliptical fits. An example of the fit is shown in Fig.\ref{fig:Slow-raster-size}(b). Figure \ref{fig:Slow-raster-size}(c) shows a linear fit between the raster current and the range of the beam distribution at the target. The x axis in Fig.\ref{fig:Slow-raster-size}(c) is the magnet current of the raster, and the y axis is the range of the beam distribution obtained from the BPMs.

\begin{figure}[tbph]
\begin{centering}
\subfloat[Location for carbon hole target in target insert]{\protect\begin{centering}
\protect\includegraphics[width=0.23\columnwidth]{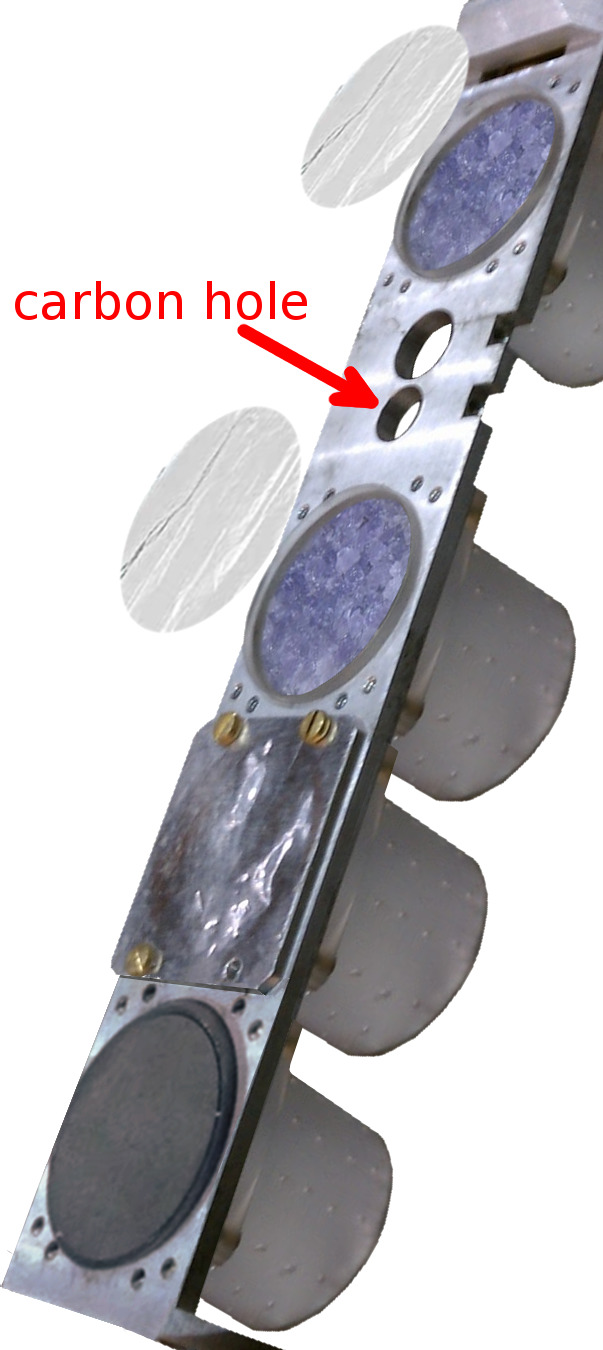}\protect
\par\end{centering}

}$\qquad$\subfloat[The shape of carbon hole in raster ADC, x and y axis are corresponding to the currents on x magnet and y magnet of slow raster, respectively.]{\protect\begin{centering}
\protect\includegraphics[width=0.58\columnwidth]{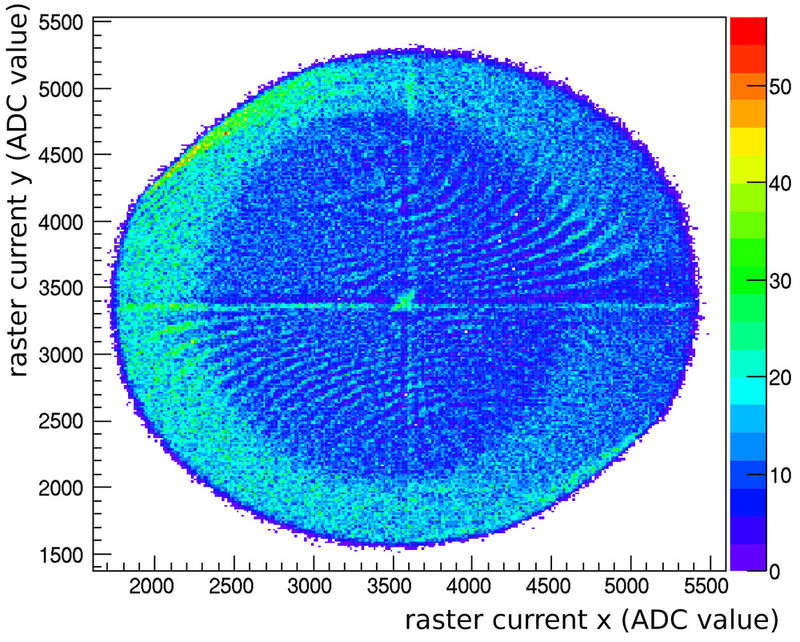}\protect
\par\end{centering}

}
\par\end{centering}

\protect\caption{\label{fig:Carbon-hole}Carbon hole method to calibrate raster}
\end{figure}

The second method for calibrating the conversion factor used a target called ``carbon hole'' as shown in Fig.\ref{fig:Carbon-hole}(a). 
 Scattered electrons were used as the trigger for recording the raster magnet current. Since the density of the target frame was much higher than that of the ``hole'', which was submerged in liquid helium, the density of events triggered from the target frame was much higher than that of the hole itself. Recorded values reveal a hole shape as shown in Fig.\ref{fig:Carbon-hole}(b). The size of the carbon hole was surveyed before the experiment, and a fit program was used to extract the radius of the recorded hole shape for that raster current. The conversion factor $F$ was then calculated as the ratio of the size of the carbon hole $S_{hole}$ and the radius of the hole shape $R_{hole}$ in the ADC:

\begin{equation}
F=\frac{S_{hole}}{2*R_{hole}}.\label{eq:convfactorhole}
\end{equation}

\subsubsection{Conversion factor for the fast raster}

\begin{figure}[tbph]
\begin{centering}
\includegraphics[width=0.5\columnwidth]{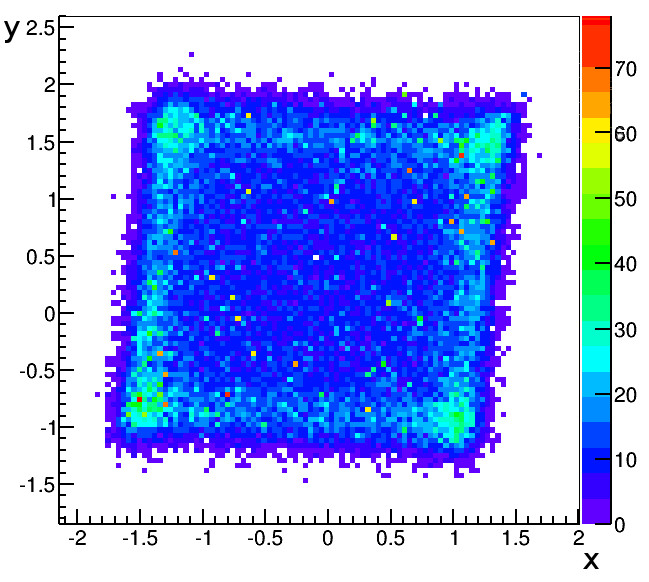} 
\par\end{centering}

\protect\caption{\label{fig:fastrasteratbpm}Beam shape formed by the fast raster at the BPM A location, the unit is millimeter}
\end{figure}

The conversion for the fast raster was the same as for the slow raster. The low pass filter for the BPM was set to a higher value than the frequency of the fast raster to see the beam shape at the BPM formed by the fast raster. For a higher frequency filter, a larger beam current was needed to get a clear pattern. The beam current chosen for calibrating the fast raster was near 300 nA, which was the safety limit for the target. The beam shape formed by the fast raster is shown in Fig.\ref{fig:fastrasteratbpm}. 

\section{Uncertainty}

The uncertainty of the final beam position at the target for each event contains several contributions: 
\begin{itemize}
\item The first part comes from the uncertainty of the calibration constant. It includes the BPM resolution for the DAQ runs used for the calibration, the uncertainty of the harp data corresponding to each calibration, and the survey uncertainties for the BPMs and harps. It contributes about 0.7 mm for the uncertainty of the position and 0.7 mrad for the uncertainty of the angle. 
\item The uncertainty on the pedestal is the largest uncertainty for the beam position measurement, contributing about 0.7$\sim$1.5 mm to the uncertainty of the position and 0.7$\sim$1.5 mrad to the uncertainty of the angle. 
\item The uncertainties from the BPM survey need to be included, since the production data and the calibration data were taken at different beamline settings when the equipment was moved. They contribute 0.5 mm to the uncertainty of the position. 
\item The uncertainty from the magnetic field map of the target was considered for the settings with the target magnet field. 
\item The uncertainties due to the size conversion of the rasters were also included. 
\end{itemize}
The position uncertainty was magnified by a factor of 5 at the target because of the short distance between the two BPMs. For example, in the straight through setting, if the uncertainty at BPM A is 0.2 mm, and at BPM B is 0.27 mm, the uncertainty at the target is 1.1 mm for position and 1.3 mrad for angle. The uncertainty for the position at the target was around 1$\sim$2 mm, while the uncertainty for the angle was 1$\sim$2 mrad.

\section{Summary}

JLab g2p experiment used a transversely polarized $NH_{3}$ target for the first time in Hall A. It put a limit of below 100 nA on the electron beam current and required a slow raster to spread beam to a large area. Two chicane magnets were used to compensate the strong transverse magnetic field. Beam-line equipment, including the BPMs, harps and associated readout system, were upgraded to allow precision measurements of the beam position at low current (50-100 nA). A software filter was used to reduce noise of the BPMs. A correction equation was used to compensate the non-linearity caused by the $\Delta/\Sigma$ equation. The harp data and the linear fit between the bpm signal and the beam current were used to extract the calibration constant of the BPM. To account for the strong target magnetic field effect, transport functions were fitted to transport the beam position from the BPMs to the target. The beam position in the x-y plane and the angle at the target location are extracted event-by-event by combining information from the BPMs and the signals from the rasters. The performance of the new devices (BPMs, harps and slow rasters) were presented along with an analysis of systematic uncertainties.

\section*{Acknowledgments}

This work was supported by DOE contract DE-AC05-84ER40150 under which the Southeastern Universities Research Association (SURA) operates the Thomas Jefferson National Accelerator Facility, and by the National Natural Science Foundation of China (11135002, 11275083), the Natural Science Foundation of Anhui Education Committee (KJ2012B179).

\bibliography{bpm_paper}

\begin{thebibliography}{10}
\expandafter\ifx\csname url\endcsname\relax
  \def\url#1{\texttt{#1}}\fi
\expandafter\ifx\csname urlprefix\endcsname\relax\def\urlprefix{URL }\fi
\expandafter\ifx\csname href\endcsname\relax
  \def\href#1#2{#2} \def\path#1{#1}\fi

\bibitem{g2pproposal}
{A. Camsonne, J. P. Chen, D. Crabb and K. Slifer, spokesperson}, {JLab E08-027
  (g2p) experiment}.

\bibitem{Crabb67}
D.~G. Crabb, W.~Meyer, Solid polarized targets for nuclear and particle physics
  experiments, Annu. Rev. Nucl. Part. Sci. 47 (1997) 67--109.

\bibitem{ADC18bob}
R.~Michaels,
  \href{http://hallaweb.jlab.org/parity/prex/adc18/prex_adc18_spec.ps}{{Precision
  Integrating HAPPEX ADC}}, JLab Technical report (unpublished).
\newline\urlprefix\url{http://hallaweb.jlab.org/parity/prex/adc18/prex_adc18_spec.ps}

\bibitem{mussonece652paper}
J.~Musson, {Functional Description of Algorithms Used in Digital Receivers},
  JLab Technical report No. JLAB-TN-14-028.

\bibitem{yan261}
{C. Yan and et al.}, {Superharp - A wire scanner with absolute position readout
  for beam energy measurement at CEBAF}, Nuclear Instruments and Methods in
  Physics Research A 365 (1995) 261--267.

\bibitem{Pierce201454}
J.~Pierce, J.~Maxwell, T.~Badman, J.~Brock, C.~Carlin, D.~Crabb, D.~Day,
  C.~Keith, N.~Kvaltine, D.~Meekins, J.~Mulholland, J.~Shields, K.~Slifer,
  \href{http://www.sciencedirect.com/science/article/pii/S0168900213016999}{Dynamically
  polarized target for the and experiments at jefferson lab}, Nuclear
  Instruments and Methods in Physics Research Section A: Accelerators,
  Spectrometers, Detectors and Associated Equipment 738~(0) (2014) 54 -- 60.
\newblock \href
  {http://dx.doi.org/http://dx.doi.org/10.1016/j.nima.2013.12.016}
  {\path{doi:http://dx.doi.org/10.1016/j.nima.2013.12.016}}.
\newline\urlprefix\url{http://www.sciencedirect.com/science/article/pii/S0168900213016999}

\bibitem{HCraster200501}
C.~Yan, \href{https://www.jlab.org/Hall-C/talks/01_06_05/yan.pdf}{{Hall C
  Polarized Target Raster System Upgrade}}, JLab Technical report
  (unpublished).
\newline\urlprefix\url{https://www.jlab.org/Hall-C/talks/01_06_05/yan.pdf}

\bibitem{Barry301}
W.~Barry,
  \href{http://www.sciencedirect.com/science/article/pii/016890029190004A}{A
  general analysis of thin wire pickups for high frequency beam position
  monitors}, Nuclear Instruments and Methods in Physics Research Section A:
  Accelerators, Spectrometers, Detectors and Associated Equipment 301~(3)
  (1991) 407 -- 416.
\newblock \href
  {http://dx.doi.org/http://dx.doi.org/10.1016/0168-9002(91)90004-A}
  {\path{doi:http://dx.doi.org/10.1016/0168-9002(91)90004-A}}.
\newline\urlprefix\url{http://www.sciencedirect.com/science/article/pii/016890029190004A}

\bibitem{carmanbpm}
C.R.Carman, J.~L. Pellegrin, The beam positions of the spear storage ring,
  SLAC-PUB-1227.

\bibitem{piotbpm}
P.Poit,
  \href{http://beamdocs.fnal.gov/AD-public/DocDB/ShowDocument?docid=1894}{Evaluation
  and correction of nonlinear effects in fnpl beam position monitors}, FNPL
  Technical report No.Beams-doc-1894-v1.
\newline\urlprefix\url{http://beamdocs.fnal.gov/AD-public/DocDB/ShowDocument?docid=1894}

\bibitem{winestosca}
R. Wines, private communication.

\bibitem{jiefieldmap}
J.~Liu,
  \href{http://hallaweb.jlab.org/experiment/g2p/collaborators/jie/2011_10_05_fieldmap_report/Target_Field_Map_Report.pdf}{Magnetic
  field mapping on a translation table}, JLab Technical report, E08-027
  Collaboration (unpublished).
\newline\urlprefix\url{http://hallaweb.jlab.org/experiment/g2p/collaborators/jie/2011_10_05_fieldmap_report/Target_Field_Map_Report.pdf}

\bibitem{chaofieldmap}
C.~Gu,
  \href{https://hallaweb.jlab.org/experiment/g2p/collaborators/chao/technotes/Chao_TechNote_TargetField.pdf}{Target
  field mapping and uncertainty estimation}, JLab Technical report, E08-027
  Collaboration (unpublished).
\newline\urlprefix\url{https://hallaweb.jlab.org/experiment/g2p/collaborators/chao/technotes/Chao_TechNote_TargetField.pdf}

\end{thebibliography}

\end{document}